\documentclass[aps,pra,10pt,twocolumn,preprintnumbers,amsmath,amssymb,showpacs]{revtex4}
\usepackage{graphicx}
\renewcommand\Re{\operatorname{Re}}
\renewcommand\Im{\operatorname{Im}}
\renewcommand\vec[1]{\mathbf #1}
\DeclareMathOperator{\STr}{STr}
\DeclareMathOperator{\sinc}{sinc}
\def\ket#1{\lvert#1\rangle}
\def\p{\vec p}

\begin{document}

\title{Excitation spectra and rf-response near the polaron-to-molecule
  transition\\ from the functional renormalization group}
\author{Richard Schmidt}
\author{Tilman Enss}

\affiliation{Physik Department, Technische Universit\"at M\"unchen,
  D-85747 Garching, Germany} 

\begin{abstract}
  A light impurity in a Fermi sea undergoes a transition from a
  polaron to a molecule for increasing interaction.  We develop a new
  method to compute the spectral functions of the polaron and molecule
  in a unified framework based on the functional renormalization group
  with full self-energy feedback.  We discuss the energy spectra and
  decay widths of the attractive and repulsive polaron branches as
  well as the molecular bound state, and confirm the scaling of the
  excited state decay rate near the transition.  The quasi-particle
  weight of the polaron shifts from the attractive to the repulsive
  branch across the transition, while the molecular bound state has a
  very small residue characteristic for a composite particle.  We
  propose an experimental procedure to measure the repulsive branch in
  a $^6$Li Fermi gas using rf-spectroscopy and calculate the
  corresponding spectra.
\end{abstract}

\pacs{11.10.Gh, 32.30.Bv, 32.70.Jz, 67.85.Lm}

\maketitle

\section{Introduction}

A single impurity $\downarrow$-atom immersed in a background of
$\uparrow$-fermions is a screened Fermi polaron \cite{landau1933}
below a critical interaction strength.  For larger interaction the
ground state changes its character and the impurity forms a molecular
bound state with the bath atoms.  This qualitative change marks the
polaron-to-molecule transition and was predicted by Prokof'ev and
Svistunov \cite{prokofev2008}.  The transition
can be regarded as the modification of the two-body problem due to
medium effects: For two non-relativistic particles with an interaction
characterized by the $s$-wave scattering length $a$, a weakly bound
molecule is the ground state for positive $a$, whereas for negative
$a$ the molecular state ceases to exist and the ground state is given
by the free atoms.

Related impurity models have been studied for many years, in
particular the Kondo effect for a fixed impurity with discrete energy
levels immersed in a Fermi sea of conduction electrons
\cite{hewson1997}.  A mobile but very heavy impurity loses its
quasi-particle character in low dimensions $d=1$, related to the
orthogonality catastrophe \cite{rosch1995}.  The transition of the
light mobile impurity poses a new and challenging many-body problem.

The polaron-to-molecule transition is observable with ultracold atoms
where the scattering length can be tuned via Feshbach resonances
\cite{bloch2008}.  Using radio-frequency (rf) spectroscopy the
lineshape, ground-state energy, and polaron quasi-particle weight have
been measured across the transition \cite{schirotzek2009}.  The
ground-state properties near the transition have been calculated using
variational wave-functions \cite{chevy2006, combescot2007,
  combescot2008, punk2009, mora2009, combescot2009, ku2009},
non-self-consistent T-matrix approximations \cite{combescot2007},
$1/N$ expansions \cite{nikolic2007}, Wilsonian renormalization group
\cite{gubbels2008}, variational Monte Carlo \cite{lobo2006} and
diagrammatic Monte Carlo (diagMC) \cite{prokofev2008}.

It turns out that even beyond the critical interaction strength the
polaron remains as a long-lived excitation above the molecular ground
state, and conversely the molecule becomes an excited state on the
polaronic side of the transition.  The proper description of these
excited states and in particular their finite lifetime remains a
difficult problem.  Using a phenomenological model in a three-loop
calculation, Bruun and Massignan have shown that the decay rates of
the excited states decrease rapidly as $\Delta\omega^{9/2}$ toward the
transition, where $\Delta\omega$ is the energy difference between the
ground and excited states \cite{bruun2010}.

Recently, interest has focused on an additional feature present in the
polaron-to-molecule transition: The renormalization of the
$\downarrow$-spectral function by the strong interactions leads to the
appearance of an additional quasi-particle excitation for positive
energies.  This excitation corresponds to a Fermi polaron interacting
repulsively with the $\uparrow$-Fermi sea.  The repulsive polaron has
a finite lifetime which becomes very small at unitarity, $a\to\infty$.
In the opposite limit of weak coupling $a\to 0$, the repulsive polaron
is long-lived and its spectral weight approaches unity.  This
justifies perturbative methods which neglect the presence of the
molecular channel \cite{bishop1973, lee1957}.  The repulsive polaron
has been studied theoretically in the context of ultracold gases by
Cui and Zhai \cite{cui2010} and more recently by Massignan and Bruun
\cite{massignan2011}.  Experimentally, the repulsive polaron has not
yet been observed in the strong-coupling regime.  However, it has
important implications for the stability of a ferromagnetic phase in
Fermi gases \cite{jo2009, conduit2009, cui2010, pilati2010,
  pekker2011, barth2011}.  Furthermore, a single $\downarrow$-fermion
in an $\uparrow$-Fermi sea corresponds to a strongly imbalanced,
two-component Fermi gas near full polarization, and the
polaron-to-molecule transition sheds light on a region of the
zero-temperature phase diagram of the polarized Fermi gas
\cite{bulgac2007, gubbels2008, punk2009}.

So far, there is a lack of theoretical work which describes all of
these features within one unified approach.  In this paper we present
a method capable of doing this: Using a novel numerical implementation
of the functional renormalization group (fRG) we are able to determine
both the full spectral functions and the quasi-particle features of
the polaron-to-molecule transition.

With the renormalization group one usually limits oneself to a few
running couplings.  In this work we develop a computational tool which
allows us to keep track of the renormalization group flow of fully
momentum and frequency dependent Matsubara Green's functions based on
an exact renormalization group equation \cite{wetterich1993}.  By
introducing an auxiliary bosonic field which mediates the interaction
between $\uparrow$- and $\downarrow$-atoms, we are able to accurately
capture the momentum and frequency dependence of both vertex functions
and propagators.  In particular for non-universal quantities our
approach, which complements the proposal by Blaizot, M\'endez-Galain,
and Wschebor \cite{blaizot2006}, may prove vital, and we demonstrate
its efficiency in the polaron problem.

Our main results from the fRG are the following.  We give the spectral
functions of the $\downarrow$-fermion and---for the first time---also
of the molecule in a wide range of interaction values.  From this we
extract the quasi-particle properties including energies, decay rates
and residues for both the polaron and the molecule.  We connect the
quasi-particle weight of the molecule with a measure of its
compositeness introduced by Weinberg \cite{weinberg1965}.  We propose
a new experimental procedure to measure the repulsive polaron in a
$^6$Li Fermi gas using rf-spectroscopy, and we employ our spectral
functions to predict the expected rf-response.

The paper is organized as follows: in section~\ref{sec:frg} we
introduce the model of the polaron-to-molecule transition and derive
appropriate fRG flow equations.  These equations are solved in section
\ref{sec:solution} for the full frequency and momentum dependent
spectral functions, which are presented and interpreted in
section~\ref{sec:spectrum}.  Section~\ref{sec:rf} is devoted to
rf-spectroscopy, and we conclude with a discussion in
section~\ref{sec:concl}.  In the appendix the fRG flow equations are
solved using the simpler derivative expansion, which already gives
qualitatively correct results.


\section{Model and RG flow equations}
\label{sec:frg}

In this work we study a two-component Fermi gas in the limit of
extreme population imbalance at $T=0$. The microscopic action
describing the system is
\begin{equation}
  \label{fermionicaction}
  S=\int_{\vec x, \tau} \Big\{
  \sum_{\sigma=\uparrow,\downarrow}
  \psi_\sigma^*[\partial_\tau - \Delta - \mu_\sigma] \psi_\sigma
  + g\psi^*_\uparrow\psi^*_\downarrow\psi_\downarrow\psi_\uparrow\Big\}
\end{equation}
in natural units $\hbar = 2m = 1$ and with imaginary time $\tau$.  The
Grassmann-valued, fermionic fields $\psi_\uparrow$ and
$\psi_\downarrow$ denote the up and down fermions, which have equal
mass $m$.  The associated chemical potentials $\mu_\sigma$ are
adjusted such that the $\uparrow$-fermions have a finite density
$n_\uparrow = k_F^3/(6\pi^2)$ while there is only a single impurity
$\downarrow$-fermion.  The atoms interact via a contact interaction
with coupling constant $g$ which is related to the $s$-wave scattering
length $a$ via $g=8\pi a$.  The T-matrix acquires a complicated
frequency and momentum dependence in the strong-coupling limit.  It is
then convenient to perform a Hubbard-Stratonovich transformation of
the action \eqref{fermionicaction} introducing a bosonic molecule
(pairing) field $\phi\sim\psi_\downarrow\psi_\uparrow$ which mediates
the two-particle interaction $g \psi^*_\uparrow \psi^*_\downarrow
\psi_\downarrow \psi_\uparrow$.  The resulting action is given by
\begin{multline}
  \label{FBaction}
  S = \int_{\vec x, \tau}
  \Big\{\sum_{\sigma=\uparrow,\downarrow}
  \psi_\sigma^*[\partial_\tau - \Delta - \mu_\sigma]\psi_\sigma
  +\phi^*G_{\phi,\Lambda}^{-1}\phi\\
  +h(\psi_\uparrow^*\psi_\downarrow^*\phi+h.c.)\Big\}
\end{multline}
with a real Yukawa coupling $h$ for the conversion of two fermions
into a molecule.  Integrating out the bosonic field $\phi$ shows that
\eqref{FBaction} is equivalent to the single-channel model
\eqref{fermionicaction} provided that $-h^2 G_{\phi,\Lambda}=g$ and
$h\to\infty$ \cite{lurie1964, nikolic2007}.  As $h^2 \sim \Delta B$
this limit corresponds to a broad Feshbach resonance \cite{bloch2008}.

The physical properties can be accessed via Green's functions which
are derivable from generating functionals.  The one-particle
irreducible vertex functions $\Gamma^{(n)}$ are obtained from the
effective quantum action $\Gamma$, which can, for instance, be
computed perturbatively in a loop expansion.  As we are interested in
the intrinsically non-perturbative regime of fermions close to a
Feshbach resonance where the scattering length $a$ diverges we employ
a different approach.  $\Gamma$ includes quantum fluctuations on all
momentum and energy scales.  The main idea of the functional
renormalization group (fRG) is to introduce an interpolating effective
flowing action $\Gamma_k$ which includes only fluctuations on momentum
scales $q\gtrsim k$ larger than the renormalization group scale $k$.
At the UV scale $k=\Lambda\to\infty$ the effective flowing action
reduces to the microscopic action $S$ which does not include any
quantum corrections.  In the infrared limit $k\to 0$, $\Gamma_k$
equals the full quantum action $\Gamma$.

The evolution, or flow, of $\Gamma_k$ with the RG scale $k$ is given
by the exact renormalization group equation \cite{wetterich1993}
\begin{equation}
  \label{wetteq}
  \partial_k\Gamma_k=\frac{1}{2}\STr\left(
  \frac{1}{\Gamma^{(2)}_k+R_k}\partial_k R_k\right).
\end{equation} 
The supertrace symbol STr denotes a loop integration over frequency
and momentum as well as the summation over all fields and internal
degrees of freedom, with a minus sign for fermions.  $\Gamma^{(2)}_k$
is the full, field dependent inverse two-point Green's function at
scale $k$, and $R_k$ is a regulator taking care of the successive
inclusion of momentum scales.  The regulator acts by imposing a large
mass term on the modes with momenta $q\lesssim k$ lower than the
cutoff scale $k$.  $R_k$ can be chosen freely as long as
$R_{k\to0}\to0$ and $R_{k\to\infty}\to \infty$.  For further details
on the fRG we refer to the literature \cite{berges2002, salmhofer2001,
  pawlowski2007}, and to its application to the BEC-BCS crossover
\cite{birse2005}.

$\Gamma_k$ is in general a functional of the fields and contains all
possible operators of the fields allowed by the symmetries.  For this
reason its exact calculation is usually impossible and one has to rely
on approximations for $\Gamma_k$.  In this work we will use the
truncation
\begin{multline}
  \label{gentrunc}
  \Gamma_k=\int_{\p,\omega}\Big\{ \psi^*_\uparrow[-i\omega
  +\p^2-\mu_\uparrow]\psi_\uparrow+\psi^*_\downarrow
  G_{\downarrow,k}^{-1}(\omega,\p)\psi_\downarrow\\ 
  +\phi^* G_{\phi,k}^{-1}(\omega,\p)\phi\Big\}+\int_{\vec x,\tau}
  h (\psi_\uparrow^*\psi_\downarrow^*\phi+h.c.) 
\end{multline}
with Matsubara frequency $\omega$.  For the momentum and frequency
dependence of the $k$-dependent, or flowing, propagators of the
$\downarrow$-fermion $G_{\downarrow,k}$ and the boson $G_{\phi,k}$ we
will present a simple approximation in terms of a gradient expansion
in the appendix.  In the following section \ref{sec:solution} we
develop a new numerical method to solve the renormalization group flow
of the propagators as completely general functions of $\omega$ and
$\p$. This will enable us to capture decay rates and dynamic effects
which is not possible in a simple gradient expansion.

Within the truncation \eqref{gentrunc} the Yukawa coupling $h$ is not
renormalized which can be seen by a simple argument: In
Eq.~\eqref{gentrunc} we neglect a term $\psi^*_\uparrow
\psi^*_\downarrow \psi_\downarrow \psi_\uparrow$ which would be
regenerated during the flow by particle-hole fluctuations
\cite{floerchinger2008}.  Similarly, 
a term $\psi^*_\uparrow\phi^*\psi_\uparrow\phi$ for the atom-dimer
interaction is neglected. Both terms would lead to a renormalization
of $h$.  Due to their omission, however, there is no diagram
generating a flow of $h$ which is why $\partial_k h\equiv0$ and $h$
remains independent of frequency and momentum.  Furthermore, we
neglect terms $(\phi^*\phi)^{n\geq2}$ which would give higher-order
corrections to the bosonic self-energy.  The majority
$\uparrow$-atoms are renormalized only by the single impurity
$\downarrow$-atom to order $1/N_\uparrow$, hence one can neglect the
renormalization of the $\uparrow$-atoms in the thermodynamic limit,
and the chemical potential $\mu_\uparrow = \epsilon_F = k_F^2/(2m)$ is
that of a free Fermi gas (we work in units where the Fermi momentum
$k_F=1$) \cite{prokofev2008}.

\begin{figure}[tb]
  \centering
  \includegraphics[width=\linewidth]{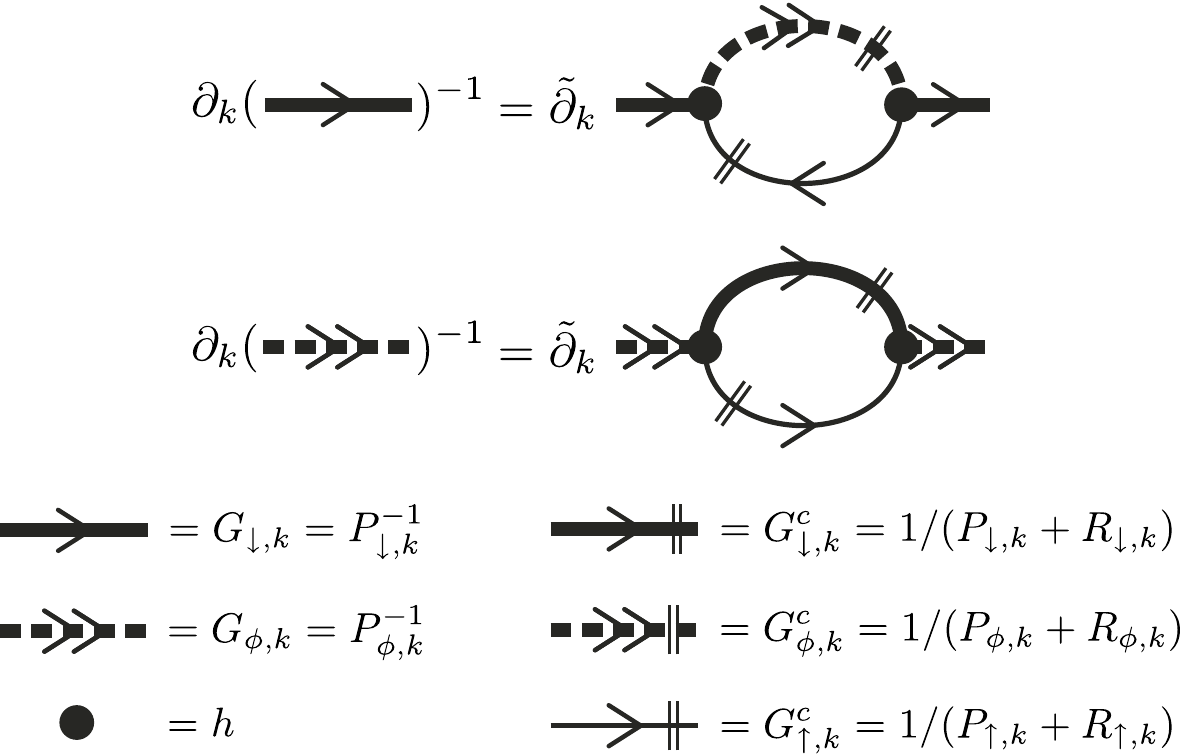}
  \caption{Diagrammatic representation of the fRG flow equations
    \eqref{genflow} for the impurity and molecule propagators.}
  \label{fig:flowequations}
\end{figure}

As Eq.~\eqref{wetteq} is a (functional) differential equation, it has
to be supplemented with appropriate initial conditions at the UV scale
$\Lambda$, which are obtained from few-body (vacuum) physics.  The
$s$-wave vacuum scattering amplitude for the interaction of an
$\uparrow$- and $\downarrow$-fermion with momenta $\vec q$, $-\vec q$
in the center-of-mass frame is given by ($q=|\vec q|$)
\begin{equation}
  \label{scattamp} 
  f(q)=\frac{1}{-1/a-iq}.
\end{equation}
$f(q)$ is related to the full molecule propagator $G_{\phi,\text
  R}^\text{vac}$ evaluated at the infrared RG scale $k=0$,
\begin{equation}
  \label{scatt_G}
  f(q)=\frac{h^2}{8\pi}\, 
  G_{\phi,\text R}^\text{vac}(\omega=2q^2,\p=0),
\end{equation}
where $\omega=2q^2$ is the total kinetic energy of the interacting
atoms.  The subscript R indicates that the analytical continuation to
the retarded function of real frequencies ($i\omega\to \omega+i0$) has
been performed.  In \cite{diehl2008} the exact vacuum molecule
propagator $G_{\phi,\text R}^\text{vac}$ has been calculated using the
fRG and agrees with the well-known result
\begin{equation}
  \label{vacprop}
  [G_{\phi,\text{R}}^\text{vac}(\omega,\p)]^{-1}
  =\frac{h^2}{8\pi}\left(-a^{-1}+\sqrt{-\frac{\omega}{2}
      +\frac{\p^2}{4}-i0}\right).
\end{equation}
This expression for $G_{\phi,\text R}^\text{vac}$ reproduces the
correct scattering amplitude \eqref{scattamp} when inserted into
Eq.~\eqref{scatt_G}, and dictates the form of the UV propagator
$G_{\phi,\Lambda}^{-1}$ for a given choice of regulator.

Furthermore, the initial condition for the fermions is given by their
form in the microscopic action \eqref{FBaction},
$G^{-1}_{\sigma,k=\Lambda}(\omega,\p)=-i\omega+\p^2-\mu_\sigma$.  As
we will see in the following, the momentum and frequency dependence of
both propagators, $G_{\phi,k}$ and $G_{\downarrow,k}$, is strongly
renormalized during the fRG flow toward the infrared, which leads to a
rich structure of the spectral functions.

After inserting the truncation \eqref{gentrunc} into the flow equation
\eqref{wetteq}, the flow of $G_{\downarrow,k}$ and $G_{\phi,k}$ is
derived by taking the appropriate functional derivatives of
Eq.~\eqref{wetteq} with respect to the fields.  One obtains the fRG
flow equations
\begin{align}
  \partial_k P_{\downarrow,k}(P)&=
  h^2\tilde\partial_k\int_Q G_{\phi,k}^c(Q) G_{\uparrow,k}^c(Q-P)\notag\\
  \partial_k P_{\phi,k}(P)&=
  -h^2\tilde\partial_k\int_Q G_{\downarrow,k}^c(Q) G_{\uparrow,k}^c(P-Q)
  \label{genflow}
\end{align}
with the multi-indices $P=(\omega,\p)$ and $Q=(\nu,\vec q)$.  The $P_k
\equiv G_k^{-1}$ on the left-hand side are the flowing inverse
propagators \emph{without} the regulator from Eq.~\eqref{gentrunc},
while the propagators $G_k^c$ on the right-hand side are regulated:
\begin{align}
  \label{eq:defP}
  G_k & \equiv 1/P_k &
  G_k^c & \equiv 1/(P_k + R_k) .
\end{align}
The tilde on $\tilde\partial_k$ indicates that the derivative with
respect to the RG scale $k$ acts only on the regulator term $R_k$ in
the cutoff propagators $G_k^c$: specifically, the single-scale
propagators read $\tilde\partial_k G_k^c = -(G_k^c)^2 \partial_kR_k$
in Eq.~\eqref{genflow}.  Note that the flow equations \eqref{genflow},
which are depicted in Fig.~\ref{fig:flowequations}, have a one-loop
structure but contain the \textit{full} propagators at scale $k$: by
integrating the flow, diagrams of arbitrarily high loop order are
generated and constantly fed back into each other. It is especially
for the latter reason that our approach goes beyond other
approximations used for the description of the polaron problem such as
for example the non-self-consistent T-matrix approximation
\cite{combescot2007, massignan2011}. The goal of this paper is to
solve the system of differential flow equations \eqref{genflow}.

In the following we choose sharp cutoff functions $R_k$ which strictly
cut off all momentum modes with $|\p|<k$ while the frequencies are not
restricted \footnote{While the solution of the un-truncated flow
  equation \eqref{wetteq} does not depend on the particular choice
  of the cutoff function, the results from the truncated flow become
  cutoff dependent.  We have implemented a class of smooth cutoff
  functions interpolating between a $k^2$ and the sharp cutoff and
  found minimal sensitivity \cite{litim2000, pawlowski2007} and also
  best agreement with Monte Carlo data for the limit of the sharp
  cutoff.}.  Then the regulated Green's functions $G_k^c$ take the
particularly simple form (for unoccupied $\downarrow$-atoms)
\begin{align}
  \label{sharpregulatorprops}
  G_{\downarrow,k}^c(\omega,\p) &=
  \frac{\theta(|\p|-k)}
  {P_{\downarrow,k}(\omega,\p)},\\ 
  G_{\phi,k}^c(\omega,\p) &=
  \frac{\theta(|\p|-k)}
  {P_{\phi,k}(\omega,\p)},\notag\\ 
  G_{\uparrow,k}^c(\omega,\p) &=
  \frac{\theta(|\p^2-\mu_\uparrow|-k^2)}
  {P_{\uparrow,k}(\omega,\p)}.\notag
\end{align}
For the $\uparrow$-atoms it is crucial to regularize the low-energy
modes around the Fermi energy $\mu_\uparrow$.  Using the Dyson
equation
\begin{equation*}
  P_k(\omega,\p)
  =G_k^{-1}(\omega,\p)
  =G^{-1}_{0}(\omega,\p)-\Sigma_k(\omega,\p),
\end{equation*}
where $G_{0}=G_{k=\Lambda}$ denotes the free (UV) and $G_k$ the full
Green's function at scale $k$, we can in particular identify the gap
term
\begin{equation}
  \label{massterm}
  m_{\downarrow,k}^2 := P_{\downarrow,k}(0,\vec 0)
  =-\mu_\downarrow-\Sigma_{\downarrow,k}(0,\vec 0).
\end{equation}
The chemical potential $\mu_\downarrow$ is the energy required to add
one $\downarrow$-atom to the system,
\begin{equation}
  \label{mudown}
  \mu_\downarrow=E(N_\downarrow)-E(N_\downarrow-1),
\end{equation}
and is independent of the cutoff scale $k$.  The interaction effects
on the $\downarrow$-fermion, which are successively included during
the flow, are captured by the $k$-dependent self-energy
$\Sigma_k(\omega,\p)$. In the polaron problem we are interested in a
two-component Fermi gas in the limit of extreme population imbalance
where one considers only a single $\downarrow$-atom, $N_\downarrow=1$,
and relation \eqref{mudown} is used to determine the ground-state
energy of the system.  This value of the chemical potential
$\mu_\downarrow$ then marks the phase transition from a degenerate,
fully polarized $\uparrow$-Fermi gas to a phase of finite
$\downarrow$-fermion density \cite{punk2009}.  Accordingly, for all
choices of $\mu_\downarrow'\leq\mu_\downarrow$ there has to be a
vanishing occupation of both $\downarrow$-fermions and molecules at
every RG scale $k$, which leads to the condition
\begin{equation}
  P_{\downarrow,k}(0,\vec 0,\mu_\downarrow')\geq0, \quad
  P_{\phi,k}(0,\vec 0,\mu_\downarrow')\geq0\quad 
  \forall \mu_\downarrow'\leq\mu_\downarrow .
\end{equation}
\begin{table}[b!]
  \begin{tabular}{|c|c|c|}
    \hline
    coupling & $(k_Fa)^{-1} < (k_Fa_c)^{-1}$
    & $(k_Fa)^{-1} > (k_Fa_c)^{-1}$ \\
    \hline
    ground state & polaron & molecule \\
    $\downarrow$ gap & $P_\downarrow(0,\vec 0)=m_\downarrow^2=0$
    & $P_\downarrow(0,\vec 0)=m_\downarrow^2>0$ \\
    $\phi$ gap & $P_\phi(0,\vec 0)=m_\phi^2>0$
    & $P_\phi(0,\vec 0)=m_\phi^2=0$ \\
    \hline
  \end{tabular}
  \caption{Conditions for the polaron and molecule ground states.}
  \label{tab:ground}
\end{table}
In order to have only a single $\downarrow$-atom or molecule,
$\mu_\downarrow$ has to be determined self-consistently such either
the $\downarrow$-atom or the molecule $\phi$ is gapless in the
infrared, $P_{\downarrow/\phi,k=0} (0,\vec 0)=0$ (ground state).  We
find that depending on the value of the dimensionless coupling $(k_F
a)^{-1}$, either the polaron or the molecule becomes the ground state,
see Table~\ref{tab:ground}.  The polaron-to-molecule transition occurs
at the critical interaction strength $(k_Fa_c)^{-1}$ at which
$m_\phi^2 = m_\downarrow^2 = 0$.


\section{RG for full spectral functions}
\label{sec:solution}

The main goal of this work is to solve the system of flow equations
\eqref{genflow} without imposing any constraints on the frequency and
momentum dependence of the polaron and molecule propagators.  This
problem can only be solved numerically, and the inverse, flowing
Green's functions $P_{\downarrow/\phi,k}(\omega,\p)$ are evaluated on
a discrete grid in frequency and momentum space,
\begin{align}
  P_{\downarrow,k}(\omega,\p)
  &\to P_{\downarrow,k}(\omega_i,p_j)=P^{ij}_{\downarrow,k}\notag\\
  P_{\phi,k}(\omega,\p)
  &\to P_{\phi,k}(\omega_i,p_j)=P^{ij}_{\phi,k}.
  \label{projection}
\end{align}
We choose a logarithmically spaced, finite grid with $\omega_i\in
(0,\ldots, \omega_\text{max})$ and $p_j\in (0,\ldots, p_\text{max})$.
As a result of rotational invariance the propagators depend only on
the magnitude of spatial momentum $p=|\p|$, and due to the condition
$P^*(\omega)=P(-\omega)$ for Euclidean (Matsubara) propagators it is
sufficient to consider positive frequencies only \cite{abrikosov1975}.
The full $(\omega,p)$ dependence is reconstructed from the finite
number of couplings $P^{ij}_{\downarrow/\phi,k}$ by cubic spline
interpolation,
\begin{align}
  P_{\downarrow/\phi,k}(\omega,p)
  &=\text{Spline}(\{P_{\downarrow/\phi,k}^{ij}\})\notag\\
  &=\sum_{\xi,\vartheta=0}^3 
  c^{ij,\xi\vartheta}_{\downarrow/\phi,k}
  (\omega-\omega_i)^\xi (p-p_j)^\vartheta,
\end{align}
with $\omega\in (\omega_i,\omega_{i+1})$, $p\in (p_j,p_{j+1})$ and
$c^{ij,\xi\vartheta}_{\downarrow/\phi,k}$ the corresponding spline
coefficients.  For the asymptotics of the propagators for high
frequency and momentum modes, $\omega>\omega_\text{max}$ and/or
$p>p_\text{max}$, we choose the simple fit models
\begin{align}
  \label{asymptotics}
  P^>_{\downarrow,k}(\omega,\p)
  &=-i\omega + \p^2 - \mu_\downarrow\\
  P^>_{\phi,k}(\omega,\p)
  &=\frac{h^2}{8\pi}\left(-a^{-1}
    +\sqrt{-\frac{i\omega}{2}+\frac{\p^2}{4}+f_{\phi,k}}\right)
  \notag
\end{align}
with $f_{\phi,k}$ determined by a continuity condition from the
numerical value of $P_{\phi,k}$ for the largest momenta $|\p| =
p_\text{max}$.

In order to keep the numerical cost of computing the flow equations
\eqref{genflow} low it is advantageous to employ the sharp momentum
regulator functions $R_{\downarrow,k}$, $R_{\uparrow,k}$, and
$R_{\phi,k}$ defined by Eq.~\eqref{sharpregulatorprops}.  This reduces
the number of loop integrations by one.  The flow equations evaluated
by our algorithm are then given by
\begin{align}
  \partial_k P_{\downarrow,k}(\omega,\p)
  &=-\frac{h^2}{(2\pi)^3} 
  \int_{-1}^{1} dx \int_{-\infty}^\infty d\nu\int_0^\infty q^2\,dq \notag\\
  &\quad\times\frac{\chi_k(p,q,x)}{P_{\phi,k}(\nu,\vec q)
    P_{\uparrow,k}(\nu-\omega,\vec q-\p)} \notag\\
  \partial_k P_{\phi,k}(\omega,\p)
  &=\frac{h^2}{(2\pi)^3} 
  \int_{-1}^{1} dx \int_{-\infty}^\infty d\nu \int_0^\infty q^2\,dq\notag\\
  &\quad\times\frac{\chi_k(p,q,x)}{P_{\downarrow,k}(\nu,\vec q)
    P_{\uparrow,k}(\omega-\nu,\p-\vec q)}
  \label{DGLfulldep}
\end{align}
where we have defined the characteristic function
\begin{multline}
  \chi_k(p,q,x)
  =\delta(q-k)\theta(\lvert(\p-\vec q)^2
  -\mu_\uparrow\rvert-k^2)\\
  +2k\theta(q-k)\delta(\lvert(\p-\vec q)^2
  -\mu_\uparrow\rvert-k^2)
  \label{charactfunc}
\end{multline}
and $x=\cos\theta$ expresses the angle $\theta$ between the momentum
vectors $\p$ and $\vec q$ such that $\lvert\p\pm\vec q\rvert^2 =
p^2+q^2\pm 2pqx$.

The initial condition at the UV scale $k=\Lambda$ is determined by the
few-body calculation \eqref{vacprop}.  As we employ no approximation
for the momentum and frequency dependence of the molecule propagator
we are able to incorporate the \textit{exact} two-body scattering
amplitude \eqref{scattamp}, in contrast to the calculation in the
appendix using the derivative expansion where this is not possible.
The vacuum problem can be solved exactly using the sharp regulators
\eqref{sharpregulatorprops}, which leads to the UV molecule propagator
\begin{multline*}
  P_{\phi,\Lambda}(\omega,\p)
  =-\frac{h^2}{8\pi a} + \frac{h^2\Lambda}{4\pi^2} \\
  -\frac{h^2}{2} \int_{\vec q} \left[
  \frac{\theta(\lvert\vec q-\tfrac{\p}{2}\rvert-\Lambda)\,
    \theta(\lvert\vec q+\tfrac{\p}{2}\rvert-\Lambda)}
  {q^2+\bigl(-\frac{i\omega}{2}+\frac{\p^2}{4}-\mu_\text{vac}\bigr)}
  - \frac{\theta(q-\Lambda)}{q^2} \right].
\end{multline*}

At each RG step we first perform the $q$ integration in
Eq.~\eqref{DGLfulldep} which is trivial due to the $\delta$-functions
in the characteristic function $\chi_k$. Next the frequency
integration in Eq.~\eqref{DGLfulldep} is carried out.  The
computational speed is greatly enhanced by mapping the numerical
integration onto an \textit{analytical} integration using the spline
polynomials in the interval $(-\omega_\text{max},\omega_\text{max})$.
$\omega_\text{max}$ is chosen such that the error in the $\omega$
integration of the outer regime $\omega>\omega_\text{max}$ introduced
due to the approximation \eqref{asymptotics} is smaller than the
accuracy of the numerical solution of the system of differential
equations \eqref{genflow}.  For the final angular integration in
$x=\cos\theta$ on the right-hand side of the flow equation
\eqref{DGLfulldep} we use a numerical integration with adaptive nodes
in order to cope with discontinuities of the integrand.

In order to obtain a stable numerical result it is sufficient to
calculate the flow of the propagators for roughly $1500$ grid points
$(\omega_i,p_j)$ in frequency and momentum space.  The corresponding
system of ordinary differential equations is straightforwardly solved
using a Runge-Kutta algorithm, which we have implemented in a version
with adaptive stepsize in RG time $t=\ln(k/\Lambda)$.  An adaptive
stepsize is essential in order to detect the kinks in the RG flow due
to the sharp Fermi surface of the $\uparrow$-fermions at zero
temperature.  We observe that about $10^4$ RG steps are necessary to
obtain an error smaller than $\epsilon \sim 10^{-5}$.

Finally, when the flow reaches the infrared, $k=0$, we end up with the
full Matsubara Green's functions $G_{\downarrow,k=0}(\omega,p)$ and
$G_{\phi,k=0}(\omega,p)$.  The initial value of $\mu_\downarrow$ for a
given $k_F a$ is adjusted such that a vanishing macroscopic occupation
of $\downarrow$-atoms and molecules is obtained at the end of the
flow, as discussed in section \ref{sec:frg}.  In order to access the
spectral functions we perform the analytical continuation to real
frequencies using a Pad\'e approximation.


\begin{figure*}
  \centering
  \includegraphics[width=\textwidth]{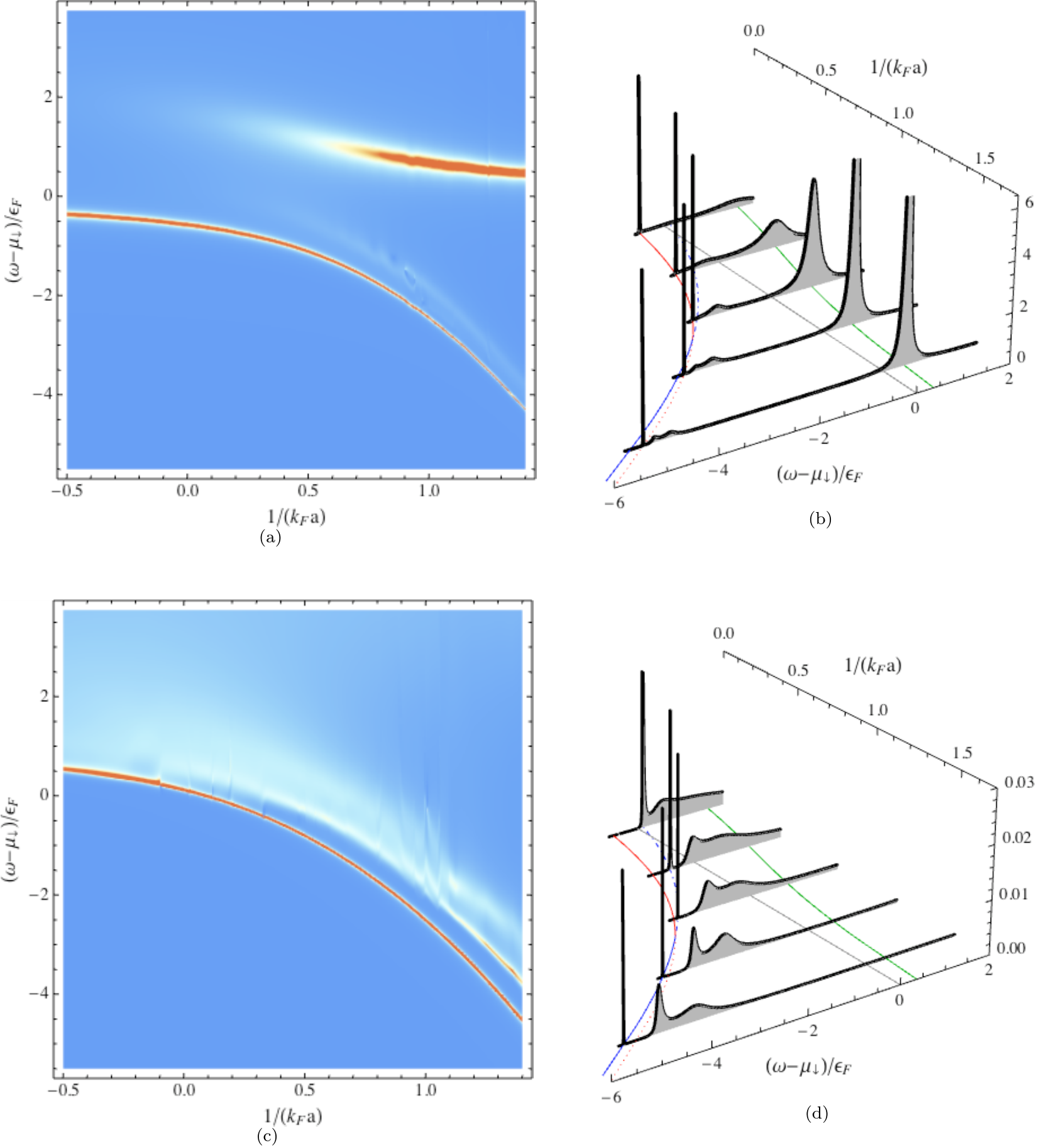}
  \caption{(color online).  Spectral functions at zero momentum in
    dependence on $1/(k_Fa)$. (a), (b): Polaron spectral function
    $A_\downarrow(\omega,\p=0)$.  (c), (d): Molecule spectral
    functions $A_\phi(\omega,\p=0)$. In order to make the
    $\delta$-function peak for the ground state visible we introduced
    an artificial width of the quasi-particles of $0.007\,\epsilon_F$.}
  \label{fig:spectralfunctions}
\end{figure*}

\section{Full spectral functions}
\label{sec:spectrum}

We now present our numerical results for the spectral functions of the
polaron and molecule across the whole transition region. First, the
Matsubara Green's functions at the end of the RG flow,
$G_{\downarrow/\phi,k=0}(\omega,\p)$, are continued analytically to
retarded Green's functions of real frequency,
$G_{\downarrow/\phi,\text R}(\omega,\p)$, using the Pad\'e
approximation.  The spectral functions are defined as
\begin{equation}
  \label{eq:specfct}
A_{\downarrow/\phi}(\omega,\p)
  = 2\Im G_{\downarrow/\phi,\text{R}}(\omega,\p) .
\end{equation}
In Fig.~\ref{fig:spectralfunctions} the zero-momentum spectral
functions $A_{\downarrow/\phi}(\omega,\p=0)$ are shown as functions of
frequency and coupling $(k_Fa)^{-1}$.

The coherent single-particle excitations at zero momentum are
determined by the solutions $\omega_\text{qp}$ of the equation
\begin{equation}
  \label{eq:quasipart}
  G_{\downarrow/\phi,\text{R}}^{-1}(\omega,\p=0)
  \Bigr\rvert_{\omega=\omega_\text{qp}} = 0
\end{equation}
for $\omega$ in the complex lower half-plane.  Near each
quasi-particle pole the retarded propagator can be approximated by the
form
\begin{equation}
  \label{eq:qpprop}
  G_{\downarrow/\phi,\text{R}}(\omega,\p=0)
  \approx \frac{Z_{\downarrow/\phi}}{\omega_\text{qp}-\omega-i0}
\end{equation}
where the real part of $\omega_\text{qp}$ determines the
quasi-particle energy
\begin{equation}
  \label{eq:Eqp}
  E_\text{qp} = \mu_\downarrow + \Re[\omega_\text{qp}].
\end{equation}
We have shifted the ground-state energy, which is zero in our
calculation (vanishing gap), to the conventional value
$\mu_\downarrow$ from Eq.~\eqref{mudown}.  The imaginary part of the
pole position determines the decay width
\begin{equation}
  \label{eq:Gammaqp}
  \Gamma_\text{qp} = -\Im[\omega_\text{qp}].
\end{equation}
A Fourier transform in time relates the decay width to the
quasi-particle lifetime
\begin{equation}
  \label{eq:tau}
  \tau_\text{qp} = \hbar/\Gamma_\text{qp}.
\end{equation}
The quasi-particle weight $Z_{\downarrow/\phi}$ is obtained from the
frequency slope at the complex pole position,
\begin{equation}
  \label{eq:Z}
  Z_{\downarrow/\phi}^{-1} = -\frac{\partial}{\partial\omega} 
  G_{\downarrow/\phi,\text R}^{-1}(\omega,\p=0)
  \Bigr\rvert_{\omega=\omega_\text{qp}}.
\end{equation}
Note that an alternative definition of the decay width in terms of the
self-energy evaluated not at the complex pole position but on the real
frequency axis,
\begin{equation}
  \label{eq:Gammaalt}
  \Gamma_\text{alt}
  = \Im \Sigma_{\downarrow/\phi,\text R}(\omega,\p=0)
  \Bigr\rvert_{\omega=\Re\omega_\text{qp}},
\end{equation}
agrees with our definition for $\Gamma_\text{qp}$ only for a single
quasi-particle pole \eqref{eq:qpprop} with $Z=1$.  However, for the
polaron problem there are further excited states and $Z<1$. Hence,
only $\Gamma_\text{qp}$ from Eq. \eqref{eq:Gammaqp} can be interpreted
as the half-width of the peaks in the spectral function and as the
inverse lifetime.

We will now in turn discuss the features seen in the spectral
functions: the peak position ($E_\text{qp}$), width
($\Gamma_\text{qp}$) and weight ($Z$), first for the polaron (upper
row of Fig.~\ref{fig:spectralfunctions}) and then for the molecule
(lower row).

\subsection{Attractive and repulsive polaron}

\begin{figure}[tb]
  \centering
  \includegraphics[width=\linewidth]{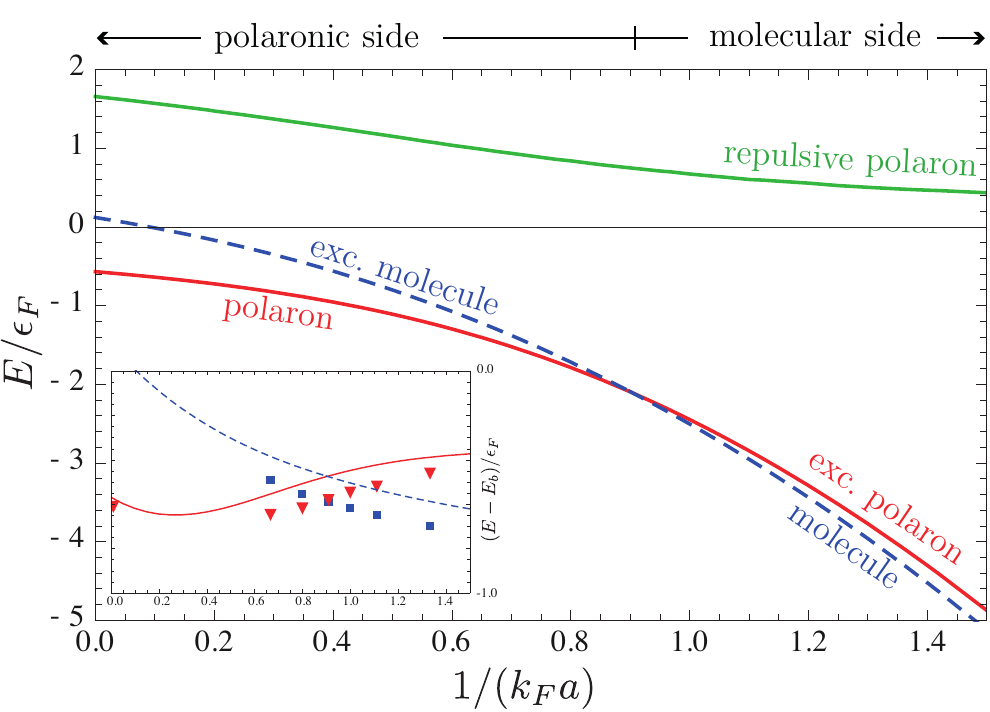}
  \caption{(color online).  Energy spectrum of the single-particle
    excitations of the polaron-to-molecule transition. In the inset we
    show our fRG result for the ground-state energy (with the
    universal dimer binding energy $E_b=-\hbar^2/(m a^2)$ subtracted)
    in comparison to the results obtained with diagMC by Prokof'ev
    and Svistunov \cite{prokofev2008} (symbols).}
  \label{fig:energyspectrum}
\end{figure}

\textbf{Energy spectrum.}  Let us first look at the energy spectrum of
the quasi-particle excitations depicted in
Fig.~\ref{fig:energyspectrum} in dependence on the coupling strength
$(k_Fa)^{-1}$.  From our data we find two coherent quasi-particle
states for the $\downarrow$-atom, the attractive and the repulsive
polaron, and one bound state for the molecule.  The attractive polaron
(red solid line) is the ground state for $(k_Fa)^{-1} < (k_Fa_c)^{-1}$
(polaronic side) but becomes an excited state for $(k_Fa)^{-1} >
(k_Fa_c)^{-1}$ (molecular side). Conversely, the molecule is the
ground state on the molecular side (blue dashed line) and an excited
state on the polaronic side, in accordance with the discussion at the
end of section \ref{sec:frg}.  For the critical coupling strength we
obtain $(k_Fa_c)^{-1} = 0.904(5)$, which agrees with the value
$(k_Fa_c)^{-1} = 0.90(2)$ obtained using diagrammatic Monte Carlo
(diagMC) by Prokof'ev and Svistunov \cite{prokofev2008}.  As shown in
the inset of Fig.~\ref{fig:energyspectrum}, also the values for the
energies agree well with diagMC (symbols).  At unitarity,
$(k_Fa)^{-1}=0$, we obtain the ground-state energy $\mu_\downarrow =
-0.57\,\epsilon_F$ while diagMC gives the value $\mu_\downarrow =
-0.615\,\epsilon_F$.  Having treated the full frequency and momentum
dependence of the propagators in the truncation \eqref{gentrunc}, we
can attribute the residual deviation in the ground state energy to the
omission of the terms $\psi_\uparrow^* \psi^*_\downarrow
\psi_\downarrow \psi_\uparrow$ and $\phi^*\psi_\sigma^*
\phi\psi_\sigma$.  The latter term describes the atom-dimer
interaction and is expected to further reduce the ground-state energy
in the transition regime in accordance with the results from the
variational wave function approach \cite{punk2009, mora2009}.  The
term $\psi_\uparrow^*\psi^*_\downarrow \psi_\downarrow\psi_\uparrow$,
generated by particle-hole fluctuations, is expected to give the main
correction in the unitarity regime \cite{combescot2008}.  Both terms
can be included in the fRG flow as additional flowing couplings, or
implicitly by using the Katanin scheme \cite{katanin2004} or
rebosonization \cite{gies2002}.

Until recently \cite{jo2009} most experiments with ultracold Fermi
gases have focused on the lower, attractive branch on the BEC side
$(k_Fa)^{-1} > 0$. There exists, however, also the repulsive polaron
branch (solid green line) which corresponds to a higher excited state
of the $\downarrow$-atom interacting \emph{repulsively} with the
$\uparrow$-Fermi sea. Our results for the energy of the repulsive
branch agree with the weak-coupling results \cite{bishop1973} for
$(k_Fa)^{-1}\gtrsim 1$. In the strong-coupling regime our energies lie
between the result from the non-self-consistent T-matrix approach
\cite{massignan2011} and the MC results for square well potentials
\cite{pilati2010}. In the polaron spectral function, cf.\
Fig.~\ref{fig:spectralfunctions}(b), one can clearly discern the
attractive polaron branch as a very sharp peak at low frequencies, and
the much broader repulsive polaron branch at higher frequencies.

\begin{figure}[tb]
  \centering
  \includegraphics[width=\linewidth]{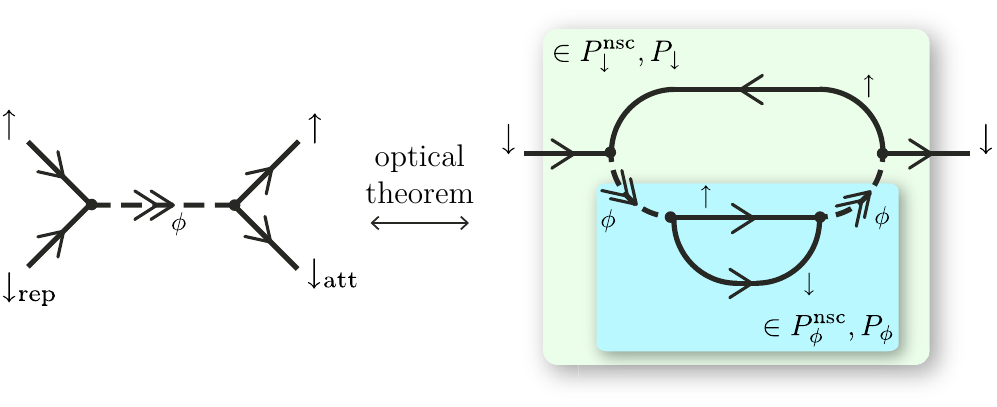}
  \caption{(color online).  Decay channel for the repulsive polaron.
    \textbf{(left)} Two-body process which leads to the decay of the
    repulsive polaron.  \textbf{(right)} Corresponding contribution to
    the $\downarrow$-atom self-energy via the optical theorem.}
  \label{fig:decayrep}
\end{figure}

\begin{figure}[tb]
  \centering
  \includegraphics[width=\linewidth]{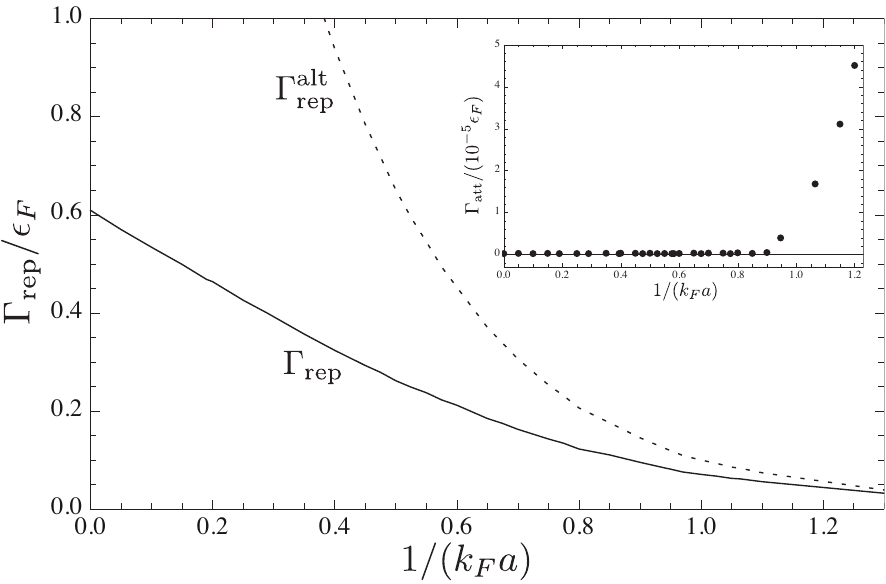}
  \caption{Decay width $\Gamma_{\text{rep}}$ of the repulsive polaron
    as a function of the coupling $(k_Fa)^{-1}$. We also show the width
    according to the approximate formula Eq.~\eqref{eq:Gammaalt}
    (dotted line). Inset: Decay width $\Gamma_\text{att}$ of the
    attractive polaron.}
  \label{fig:gammarep}
\end{figure}

\textbf{Decay widths.}  The repulsive polaron has a large decay width
$\Gamma_{\text{rep}}$, as calculated from Eq. \eqref{eq:Gammaqp} and
depicted in Fig.~\ref{fig:gammarep}, and correspondingly a short
lifetime.  The leading-order decay channel for the repulsive polaron
is the process shown in Fig.~\ref{fig:decayrep}(left) where the
repulsive polaron, which is an excited state, decays to the
attractive, and energetically lower lying, polaron due to the
interaction with an $\uparrow$-atom. This diagram can be translated
via the optical theorem into a contribution to the imaginary part of
the $\downarrow$-atom self-energy, as depicted in
Fig.~\ref{fig:decayrep}(right).  This self-energy diagram is already
included in the non-self-consistent T-matrix propagator
$P_\downarrow^\text{nsc}$, and has been studied recently using this
approximation \cite{massignan2011}.  Of course, this diagram is also
included in our fRG approach.

In the weak-coupling limit $(k_Fa)^{-1}\to\infty$ the excitation
becomes sharp, $\Gamma_{\text{rep}}\to0$, and the repulsive polaron is
a well-defined quasi-particle. Toward unitarity, $\Gamma_{\text{rep}}$
grows but remains a well-defined, finite quantity even at unitarity.
We find that indeed $\Gamma_\text{rep}$, and not the approximation
$\Gamma_\text{rep}^\text{alt}$ from Eq. \eqref{eq:Gammaalt}, is the
correct half-width at half-height of the respective peak in the
polaron spectral function in Fig.~\ref{fig:spectralfunctions}(b). For
$(k_Fa)^{-1}<0.6$ the energy $E_\text{rep}$ of the repulsive branch
exceeds the bath Fermi energy, $E_\text{rep} > \epsilon_F$.  At this
point it is energetically favorable to spin-flip the impurity atom,
which can be interpreted as the condition for the onset of saturated
ferromagnetism \cite{cui2010, barth2011}.  At the same time the decay
width $\Gamma_\text{rep}>0.2\,\epsilon_F$ is large, which potentially
destabilizes a ferromagnetic phase \cite{pekker2011}.

\begin{figure}[tb]
  \centering
  \includegraphics[width=\linewidth]{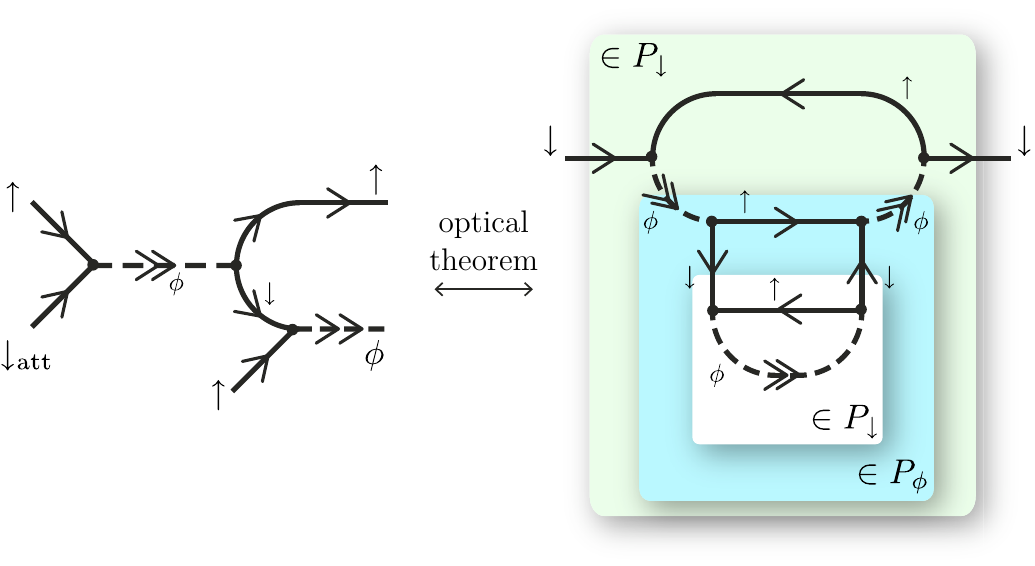}
  \caption{(color online).  Decay channel for the attractive polaron.
    \textbf{(left)} Three-body recombination process which leads to
    the decay of the attractive polaron.  \textbf{(right)} Corresponding
    contribution to the $\downarrow$-atom self-energy via the optical
    theorem (there is also a contribution with crossed lines).}
  \label{fig:decayatt}
\end{figure}

On the polaronic side $(k_Fa)^{-1} < (k_Fa_c)^{-1}$ the attractive
polaron is the stable ground state with decay width
$\Gamma_{\text{att}} = 0$, while on the molecular side it is an
excited state with finite lifetime and decay width
$\Gamma_{\text{att}} > 0$, see Fig. \ref{fig:gammarep}(inset). This
decay is much weaker and also qualitatively different from the
repulsive channel. The attractive polaron can decay by a three-body
recombination process as shown in Fig.~\ref{fig:decayatt}(left).  Via
the optical theorem this process can be translated into a contribution
to the $\downarrow$-atom self-energy as depicted in
Fig.~\ref{fig:decayatt}(right), plus an additional contribution with
crossed lines. This decay channel has recently been studied using an
explicit three-loop calculation \cite{bruun2010}.  The resulting
finite lifetime cannot be seen in the non-self-consistent T-matrix
approximation, where the self-energy corrections of the
$\downarrow$-atom---indicated by the inner white box
$P_\downarrow$---are \emph{not} fed back into the T-matrix $\sim
P_\phi^{-1}$ \cite{combescot2007, massignan2011}.  In contrast to the
non-self-consistent T-matrix calculation, our fRG includes the full
feedback of both the $\downarrow$ and $\phi$ self-energies, denoted by
bold internal lines in the flow equations in
Fig.~\ref{fig:flowequations}.  Therefore, the contributions from the
decay diagram in Fig.~\ref{fig:decayatt}(right), and many more, are
automatically included in our approach.

\begin{figure}[tb]
  \centering
  \includegraphics[width=\linewidth]{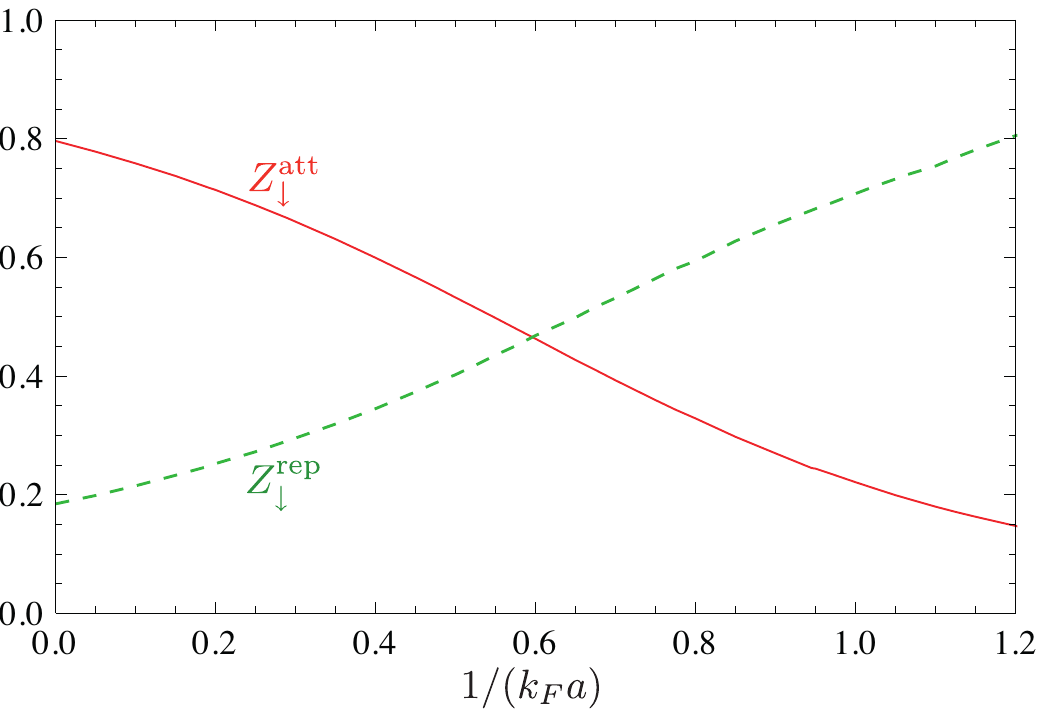}
  \caption{(color online).  Quasi-particle weight $Z_\downarrow$ of
    the attractive (solid line) and repulsive (dashed line) polaron.
    The weights of the two quasi-particle peaks in the $\downarrow$
    spectral function almost completely make up the total spectral
    weight, and the contribution from the incoherent background is
    very small.}
  \label{fig:Zpol}
\end{figure}

\textbf{Quasi-particle weights.}  Fig.~\ref{fig:Zpol} depicts the
quasi-particle weights of the attractive and repulsive polaron
computed using Eq.~\eqref{eq:Z}.  On the polaronic side the attractive
polaron state contains most of the weight, but as one moves toward the
molecular side the spectral weight gradually shifts to the repulsive
branch, and the corresponding peak in the polaron spectral function in
Fig.~\ref{fig:spectralfunctions}(b) becomes larger.  We find that the
attractive and repulsive branches almost completely make up the total
spectral weight, hence the contribution from the incoherent background
is very small.  This is also apparent in the polaron spectral function
in Fig.~\ref{fig:spectralfunctions}(b).

Our results for the quasi-particle weights agree well with those from
the non-self-consistent T-matrix and variational wave-function
approaches.  At unitarity we obtain for the attractive polaron
$Z_{\downarrow,\text{att}}= 0.796$ compared to the variational value
$Z_{\downarrow,\text{att}}= 0.78$ \cite{chevy2006, punk2009}.  For the
repulsive polaron at $(k_Fa)^{-1}=1$ we find
$Z_{\downarrow,\text{rep}} = 0.71$, in agreement with the recent
non-self-consistent T-matrix calculation \cite{massignan2011}.

Note that there is an alternative definition of the quasi-particle
weight \cite{punk2009}
\begin{equation}
  \label{eq:Zalt}
  Z_\downarrow^\text{alt} = \lim_{t\to\infty} \bigl\lvert
  G_\downarrow(t,\p=0) \bigr\rvert.
\end{equation}
This definition has to be treated with care: on the molecular side of
the transition the polaron acquires a finite decay width
$\Gamma_\downarrow > 0$ but nonetheless continues to be a well-defined
quasi-particle with finite spectral weight, as can be seen from
Fig.~\ref{fig:spectralfunctions}.  However, definition \eqref{eq:Zalt}
yields zero as soon as $\Gamma_\downarrow > 0$ and in this case cannot
be interpreted as a measure of spectral weight anymore.  In contrast,
the definition \eqref{eq:Z} remains correct for a finite decay width
and accordingly our data for $Z_\downarrow$ shows no discontinuity at
the transition. In the experiment the finite lifetime of the
attractive polaron on the molecular side complicates the direct
measurement of $Z_\downarrow$ by radio-frequency spectroscopy because
the molecular state and not the attractive polaron becomes occupied as
the initial state (cf. section \ref{sec:rf}).

\subsection{Molecule}

\textbf{Energy spectrum.}  The molecule spectral function in
Fig.~\ref{fig:spectralfunctions}(d) displays a sharp quasi-particle
peak of the bound state at low frequencies, followed by an incoherent
background at higher frequencies which actually carries most of the
spectral weight. Note that this background is not taken into account
in the simple derivative expansion in the appendix nor in the
Wilsonian RG approach \cite{gubbels2008}.  On the molecular side
$(k_Fa)^{-1} > (k_Fa_c)^{-1}$ the molecule is the ground state and is
clearly separated from the incoherent continuum.  On the polaronic
side of the transition the molecule becomes an unstable, excited state
and develops a clearly visible finite decay width in the spectral
function.

\begin{figure}[tb]
  \centering
  \includegraphics[width=\linewidth]{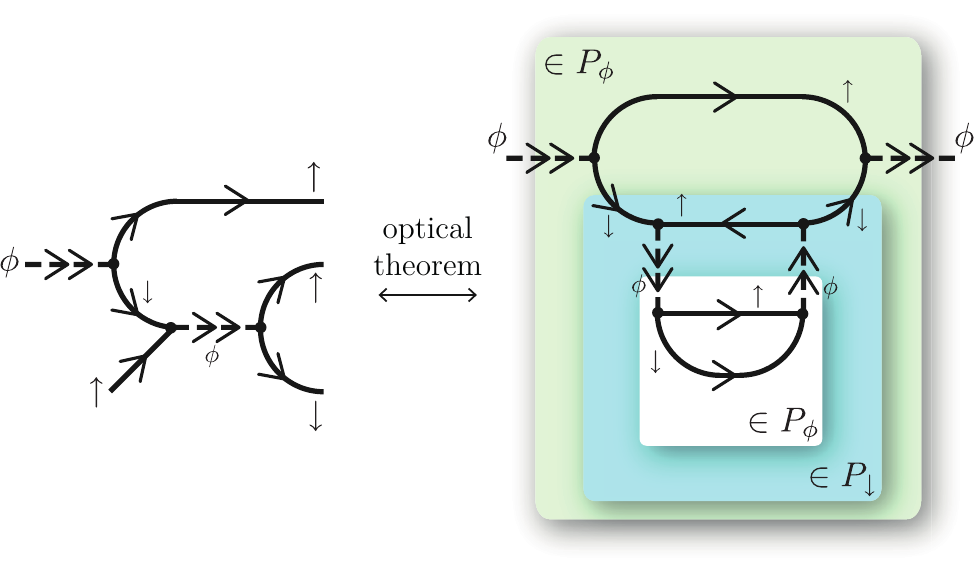}
  \caption{(color online).  Leading decay channel for the excited
    molecular state.  \textbf{(left)} Three-body recombination process
    which leads to the decay of the molecular state.  \textbf{(right)}
    Corresponding contribution to the molecule self-energy via the
    optical theorem (there is also a contribution with crossed
    lines).}
  \label{fig:decaymol}
\end{figure}

\textbf{Decay widths.}  The leading decay channel of the excited
molecule state is via the three-body recombination process shown in
Fig.~\ref{fig:decaymol}(left).  Via the optical theorem this process
can be translated into a contribution to the molecule self-energy as
depicted in Fig.~\ref{fig:decaymol}(right).  Similarly to the
attractive polaron, in the non-self-consistent T-matrix approximation
the $\downarrow$-atom self-energy corrections in $P_\downarrow$ are
\emph{not} fed back into the T-matrix $\sim P_\phi^{-1}$, and the
molecule does not decay.  In contrast, the diagram
Fig.~\ref{fig:decayatt}(right) is included in the fRG, which leads to
the visible broadening in the spectral function.

\begin{figure}[tb]
  \centering
  \includegraphics[width=\linewidth]{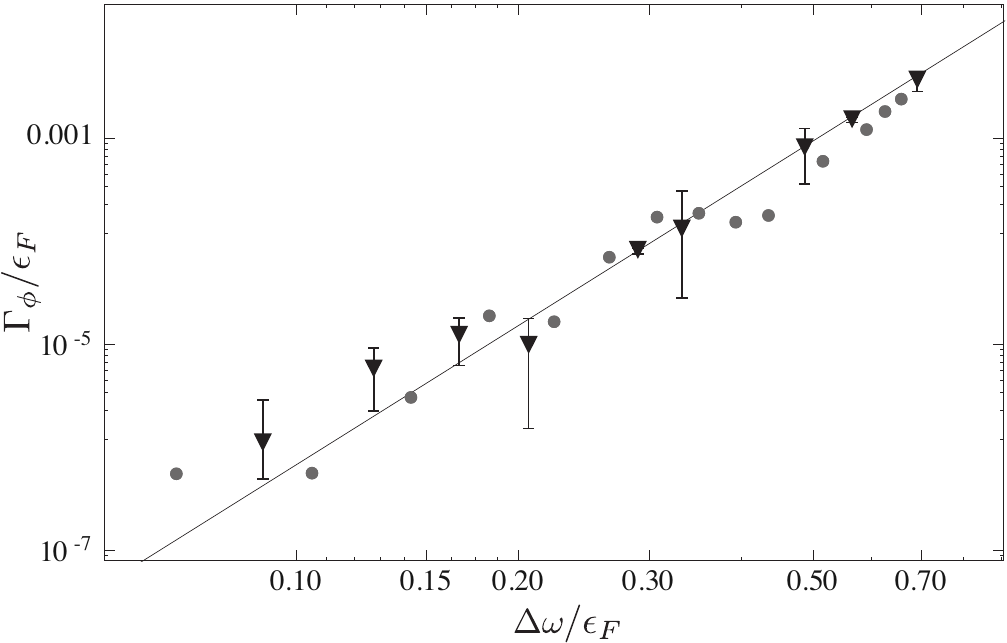}
  \caption{The decay width $\Gamma_\phi$ of the excited molecular
    state as a function of the energy difference $\Delta\omega =
    E_\phi-E_{\downarrow,\text{att}}$ between the excited molecule and
    attractive polaron ground state.  The solid line indicates the
    power-law scaling $\Gamma_\phi \propto \Delta\omega^{9/2}$.}
  \label{fig:powerlaw}
\end{figure}

Ref.~\cite{bruun2010} shows by an analytical calculation of the phase
space for three-loop diagrams of the type in Fig. \ref{fig:decaymol}
that the decay width of the molecule scales as
\begin{align}
  \label{eq:molscaling}
  \Gamma_\phi & \propto \Delta\omega^{9/2} &
  \Delta\omega & = E_\phi - E_{\downarrow,\text{att}}
\end{align}
where $\Delta\omega$ is the difference between the energy levels of
the excited molecule and the attractive polaron ground state.  In
Fig.~\ref{fig:powerlaw} we show $\Gamma_\phi$ as a function of
$\Delta\omega$ in a double logarithmic plot.  The large fluctuations
of our numerical data are due to the accuracy of the Runge-Kutta
integration as well as due to the restriction to a finite number of
Matsubara frequencies.  We have estimated the corresponding error by
comparing the results for different grids with varying number and
position of (Matsubara) frequencies. The solid line in
Fig.~\ref{fig:powerlaw} indicates the power law $\Delta\omega^{9/2}$.
The triangles in Fig.~\ref{fig:powerlaw} correspond to a calculation
with a higher number of frequencies and we find a convergence to the
solid curve for decay widths larger than our numerical integration
accuracy $\epsilon=10^{-5}$. This indicates that the error for larger
$\Gamma_\phi$ can be attributed to the Pad\'e approximation, while for
$\Gamma_\phi<\epsilon$ the accuracy of our results becomes limited by
the absolute error of our numerical integration.  With our fRG
calculation we are thus able to verify the prediction by Bruun and
Massignan \cite{bruun2010}, and the correctness of the power law
attests to the strength of our method to describe many features of the
polaron-to-molecule transition in one unified approach.

\begin{figure}[tb]
  \centering
  \includegraphics[width=\linewidth]{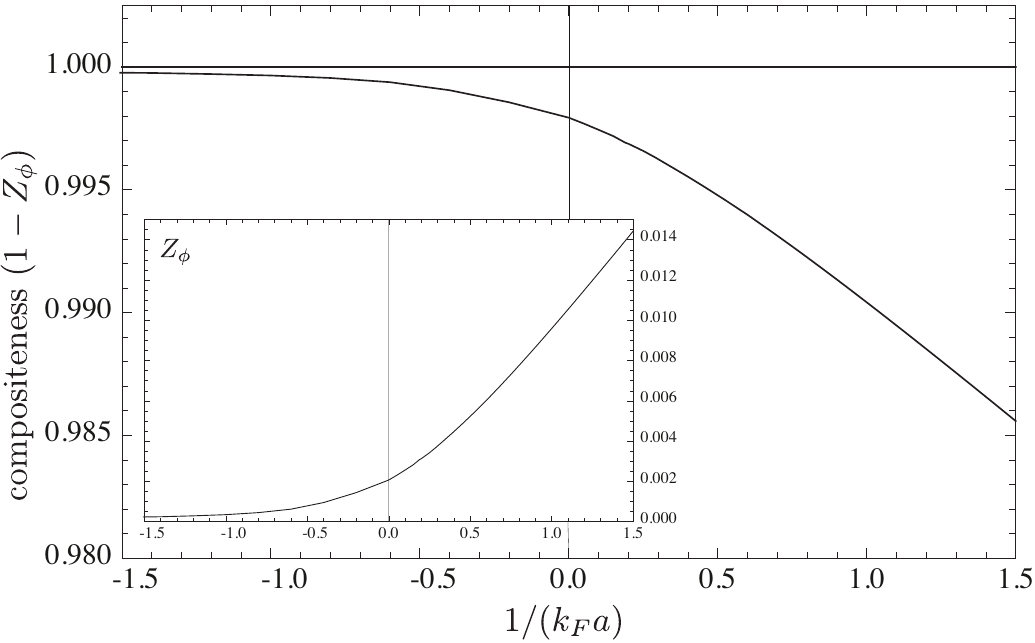}
  \caption{Compositeness $(1-Z_\phi)$ of the molecular bound state.  A
    value of $100\%$ would indicate that the molecule has no overlap with
    elementary closed-channel bosons. Inset: Molecular residue $Z_\phi$.}
  \label{fig:compositeness}
\end{figure}

\textbf{Quasi-particle weight and compositeness.}  The quasi-particle
weight $Z_\phi$ of the molecular bound state in the spectral function
in Fig.~\ref{fig:spectralfunctions}(d) is very small, $Z_\phi \approx
0.002$ at unitarity, and increases slowly toward the molecular limit,
see Fig.~\ref{fig:compositeness}(inset).  For broad resonances,
$h^2\sim\Delta B\to\infty$, the two-channel model \eqref{FBaction} is
equivalent to the single-channel model \eqref{fermionicaction} where
$Z_\phi=0$ \cite{lurie1964}. Specifically, we obtain for the weight of
the bound state in vacuum
\begin{align}
  \label{eq:Zphi}
  Z_\phi = \frac{32\pi}{h^2a} \qquad \text{(vacuum,\;$\Lambda\to\infty$)}.
\end{align}
This is consistent with the interpretation of $Z_\phi$ as the
closed-channel admixture (cf.\ Eq.~(29) in \cite{bloch2008}). In our
calculation we set the physical UV cutoff scale to $\Lambda=10^3k_F$,
which is of the order of the inverse Bohr radius, and choose
$h^2<\infty$.  We observe $Z_\phi\propto 1/(k_Fa)$ on the BEC side and a
deviation from the vacuum scaling close to unitarity, which may be due to
a combination of finite density corrections and the admixture of
closed-channel molecules in the microscopic action by choosing finite
values of $h$ and $\Lambda$.

Furthermore, the quasi-particle weight $Z$ can be interpreted as the
overlap between the ``true'' particles and the ``elementary'', or
bare, particles in the microscopic action \eqref{FBaction}.  The
attractive Fermi polaron becomes elementary,
$Z_{\downarrow,\text{att}}\to 1$, in the BCS limit ($k_Fa\to 0^-$),
while in the opposite limit of $k_Fa\to 0^+$ the repulsive Fermi
polaron becomes elementary, $Z_{\downarrow,\text{rep}}\to 1$. Near
unitarity, both excitations have a sizable weight.  In contrast, the
molecular bound state is almost exclusively a composite particle in
the whole transition region. Indeed, the deviation of the
quasi-particle weight from unity, $1-Z_\phi$, is a well-established
measure of compositeness in nuclear physics \cite{weinberg1965}, and
in Fig.~\ref{fig:compositeness} we show that for our choice of the
Yukawa coupling $h$ the compositeness of the molecule is very large
($>98\%$). This is consistent with the measurement of a small
molecular weight $Z_\phi$ for a balanced $^6$Li Fermi gas close to a
broad Feshbach resonance by Partridge \textit{et al.}
\cite{partridge2005}. In experiments with a narrow Feshbach resonance
the compositeness will decrease and a single-channel description
becomes invalid. A strength of the fRG approach is that both
situations are naturally described by tuning the values of $h$ and
$G_{\phi,\Lambda}$.

In the vacuum there exists no molecular state for negative scattering
length $a$ as can easily seen from Eq.~\eqref{vacprop}, which has no
bound-state pole. In the presence of a medium of $\uparrow$-fermions,
however, the molecule propagator develops an excited bound-state pole
also for negative scattering length $a$.  This bound state will become
the superfluid ground state if the impurity density exceeds a critical
threshold \cite{punk2009}.


\section{rf-response of the $^6$Li Fermi gas}
\label{sec:rf}

As a final application we want to connect our results for the
polaronic spectral function to experimentally observable
radio-frequency (rf) spectra. The attractive branch of the
polaron-to-molecule transition has been studied experimentally by
Schirotzek \textit{et al.}\ \cite{schirotzek2009} using a population
imbalanced, two-component mixture of $^6$Li atoms. In the experiment
the rf-response of the system has been used to infer information about
the low-frequency behavior of the fermionic spectral functions. For
instance, the ground-state energy and the residue $Z_\downarrow$ of
the $\downarrow$-fermions were measured and confirmed the theoretical
predictions.

\begin{figure}[tb]
  \centering
  \includegraphics[width=\linewidth]{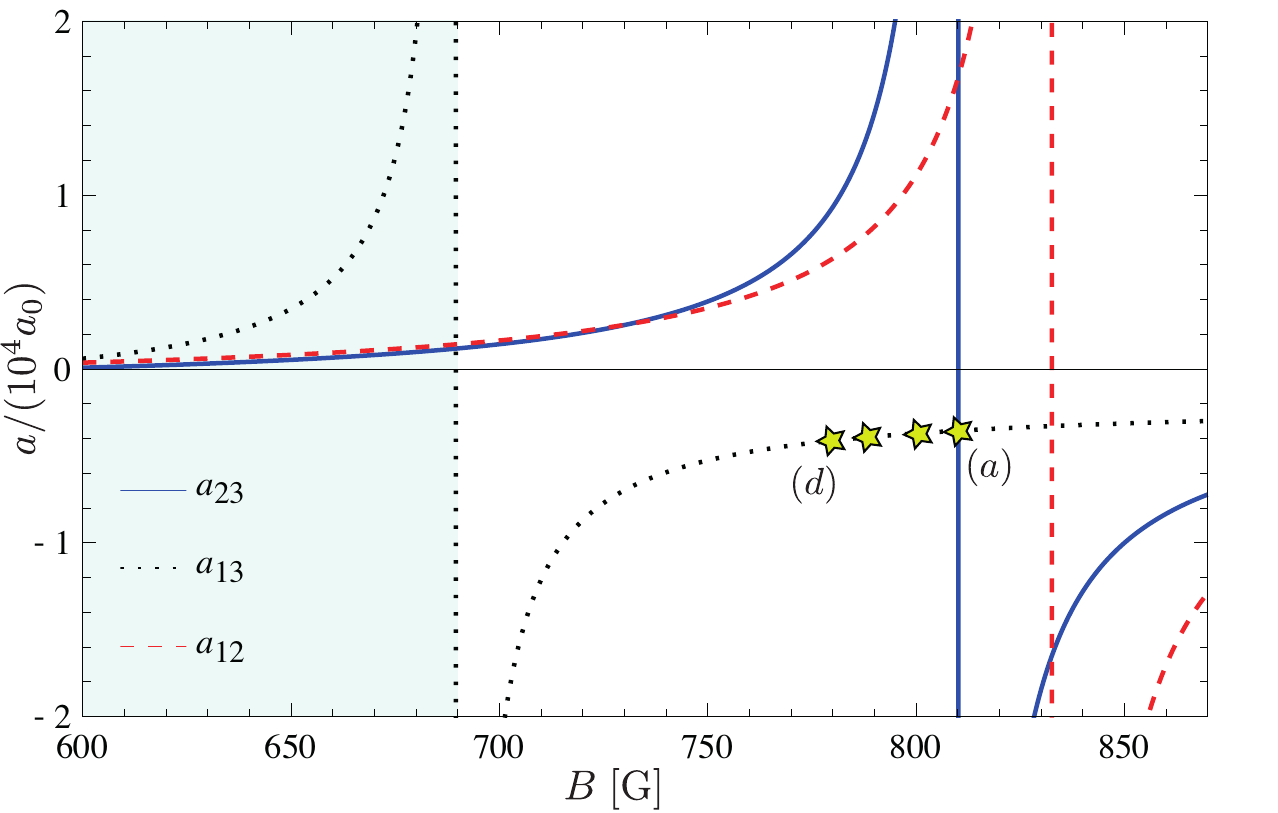}
  \caption{(color online).  Scattering length profile of $^6$Li atoms
    in the three lowest hyperfine states as calculated by P.~S.
    Julienne based on the model described in \cite{bartenstein2005}.
    The stars indicate the magnetic field values for which we
    determine the rf-response of the imbalanced Fermi gas in
    Fig.~\ref{fig:rfspectra}.}
  \label{fig:scattprofile}
\end{figure}

In order to measure the rf-response, an rf-pulse is applied to the
system which drives the transition of the atoms to a third, initially
empty state. In Fig.~\ref{fig:scattprofile} we show the scattering
length profile of $^6$Li versus the magnetic field. In the experiment
\cite{schirotzek2009} a mixture of fermions initially in the hyperfine
states $\ket{1}$ and $\ket{3}$ had been prepared in a range of the
external magnetic field $B= 630 \ldots 690\,$G (shaded area). In this
regime the scattering length $a_{13}$ in the initial state is large and
positive, while the final state scattering lengths $a_{12}$, $a_{23}$
are rather small.

Information about the spectral function can be accessed from
rf-spectroscopy, for example by populating the particle under
investigation up to the energies one is interested in and then driving
the transition to a weakly interacting final state for which the
spectral function is well-known. This route had been taken for the
study of the attractive polaron. The repulsive polaron branch, on the
other hand, has not yet been observed directly in experiments.  The
main complication is that the repulsive polaron has a very short
lifetime in the strong-coupling regime of interest (cf.\
Fig.~\ref{fig:gammarep}). Hence, its macroscopic population is
inhibited on longer time scales, and even after a fast ramp to the
desired magnetic field most minority atoms will have decayed into the
respective ground state.  A similar situation arises in the detection
of Efimov trimers in a three-component mixture of $^6$Li atoms
\cite{lompe2010, nakajima2011}.  The decay of the repulsive branch is
also of relevance for the balanced system and the competition between
ferromagnetic order and molecule formation \cite{jo2009, pekker2011}.
  
In this section we propose an experimental procedure to circumvent
these difficulties and directly analyze the spectral function of the
repulsive polaron. A strongly imbalanced two-component $^6$Li Fermi
gas is prepared in hyperfine states $\ket{1}$ and $\ket{3}$ for
magnetic fields $B>690\,$G across the $(1,3)$ Feshbach resonance. In
this regime the initial scattering length is negative, $a_{13}<0$. One
then drives an rf-transition to the final state $\ket{2}$ which is
characterized by large, positive scattering lengths $a_{12}$ and
$a_{23}$ and thus strong interactions. Because the attractive polaron
spectral function of the initial state, with its negative scattering
length $a_{13}$, is well understood both experimentally and
theoretically, the final-state spectral function can then be analyzed
in a controlled fashion.

Within linear response theory the induced transition rate from the
initial state $\ket{i}$ to the final state $\ket{f}$ is given by
\cite{punk2007, punk2009b, haussmann2009}
\begin{equation}
  I(\omega_L)=2\Omega^2 \Im\chi_\text{R}(\mu_f-\mu_i-\omega_L)
\end{equation}
where the Rabi frequency $\Omega$ is given by the coupling strength of
the rf-photon to the atomic transition, $\mu_i$ ($\mu_f$) the initial
(final) state chemical potential, and $\omega_L$ denotes the
rf-frequency offset with respect to the free rf-transition frequency.
Neglecting the momentum of the rf-photon, the retarded
rf-susceptibility $\chi_\text{R}$ is given by the analytical
continuation to real frequencies of the correlation function in
imaginary time $\tau$ (Matsubara frequency $\omega$)
\begin{multline}
  \label{rfsusc}
  \chi(\omega)=-\int_\vec r\int_{\vec r'}\int_\tau 
  e^{i \omega \tau}\langle T_\tau \psi_f^\dagger(\vec r, \tau)
  \psi_i(\vec r, \tau) \\
  \times \psi_i^\dagger(\vec r', 0)\psi_f(\vec r', 0)\rangle,
\end{multline}
where $T_\tau$ is the imaginary time-ordering operator.
Eq.~\eqref{rfsusc} leads to various diagrammatic contributions which
are in general difficult to handle if the final-state interactions are
not negligible \cite{pieri2009}. Here we will calculate
Eq.~\eqref{rfsusc} in a simple approximation with full Green's
functions but without vertex corrections.  In this approximation
Eq.~\eqref{rfsusc} yields the susceptibility in Matsubara frequency
\begin{equation}
  \chi(\omega)=\int_{\vec k, \nu} G_i(\vec k, \nu)\,
  G_f(\vec k,\nu+\omega).
\end{equation}
The rf-response in real frequency is then given by
\begin{multline}
  \label{losusc}
  I(\omega_L)=\Omega^2\int_{\vec k}
  \int_0^{\mu_i-\mu_f+\omega_L}\frac{d\nu}{2\pi}\\
  \times A_f(\vec k, \nu)\;A_i(\vec k, \nu+\mu_f-\mu_i-\omega_L)
\end{multline}
where $\mu_i-\mu_f+\omega_L>0$.

\begin{table}[tb]
  \begin{tabular}{|c|c|c|c|c|}
    \hline
    & $B$ field [G] & $(k_Fa_{13})^{-1}$ & $(k_Fa_{23})^{-1}$
    & $(k_Fa_{12})^{-1}$ \\
    \hline
    (a) & 810.3 & -1.88 &0.0& 0.39\\
    (b) & 800.8 & -1.80 & 0.2&0.58\\
    (c) & 788.2 & -1.70 &0.5 &0.86\\
    (d) & 780.6 & -1.62 & 0.7&1.04\\
    \hline
  \end{tabular}
  \caption{Interaction parameters at the four transitions indicated in
    Fig.~\ref{fig:scattprofile}, using $k_{F\uparrow}=0.00015\, a_0^{-1}$.}
  \label{tab:trans}
\end{table}

In Eq.~\eqref{losusc} the initial-state spectral function $ A_i$ is
probed for negative frequencies only. Hence, there is no rf-response
for the pure polaron problem at vanishing density and chemical
potential $\mu_\downarrow^{(0)}$. In the experiment one has, however,
a small but finite concentration $x=n_\downarrow/n_\uparrow$ of
$\downarrow$-fermions which leads to an observable rf-response.  In
order to describe the experimental situation we therefore need a
calculation for a finite minority ($\downarrow$) density characterized
by a chemical potential $\mu_i=\mu_\downarrow>\mu_\downarrow^{(0)}$.
Within the fRG framework such a calculation requires the regulator
$R_\downarrow$ in Eq.~\eqref{sharpregulatorprops} to be adjusted in
order to cope with the finite Fermi surface of $\downarrow$-fermions
which complicates the computation. Fortunately, our calculation for
the polaron problem shows that for negative scattering length, where
the polaron is the ground state and decay processes do not matter, the
fRG results are in excellent agreement with the results from a
non-self-consistent T-matrix approach. We may therefore use this
approach instead of a full-feedback fRG for the calculation of the
imbalanced Fermi gas of finite densities $n_\uparrow$ and
$n_\downarrow$ in order to determine the initial-state spectral
function. The non-self-consistent calculation was also done by Punk
and Zwerger \cite{punk2007} and is obtained in our fRG formulation by
simply switching off the feedback of the $\downarrow$-atom self-energy
into the molecule flow.

\begin{figure}[tb]
  \centering
  \includegraphics[width=\linewidth]{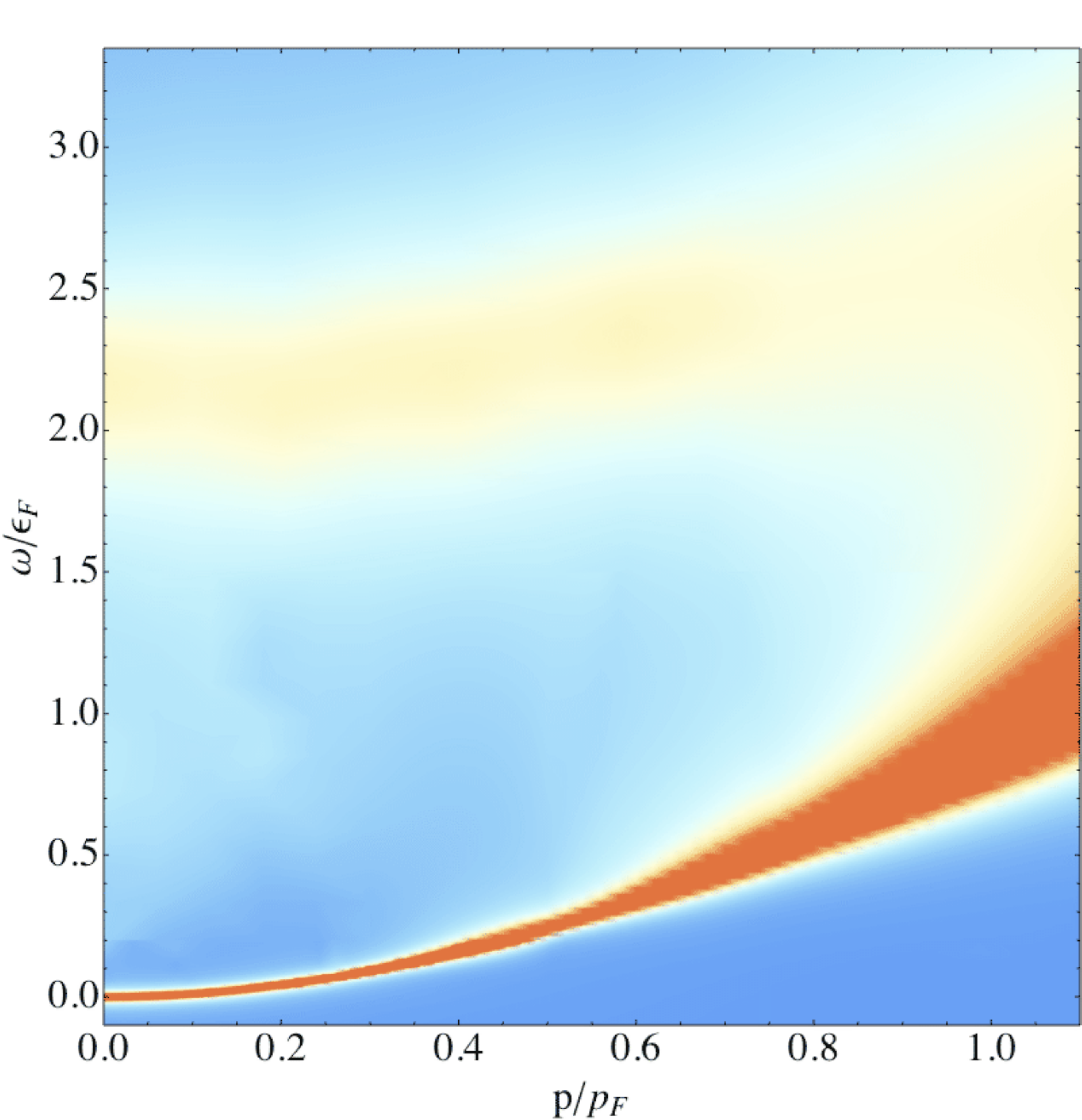}
  \caption{(color online).  Full momentum and frequency dependence of
    the polaron spectral function $A_\downarrow(\omega,\p)$ at
    unitarity $(k_Fa)^{-1}=0$.}
  \label{fig:specmom}
\end{figure}

Because the occupation of $\downarrow$-atoms is small only the
low-momentum modes are relevant. The $\downarrow$-atoms form a
degenerate Fermi gas of polaronic quasi-particles and the spectral
function can be approximated by \cite{lobo2006}
\begin{equation}
  \label{initialA}
  A_i(\omega,\p)
  =2\pi Z_i\, \delta\Bigl(\omega-\frac{\p^2}{2m_\downarrow^*}+\Delta\Bigr),
\end{equation}
where $Z_i$ is the residue and $m_\downarrow^*$ is the effective mass
of the impurity atoms.  $\Delta$ determines the impurity concentration
$x$.

\begin{figure}[tb]
  \centering
  \includegraphics[width=\linewidth]{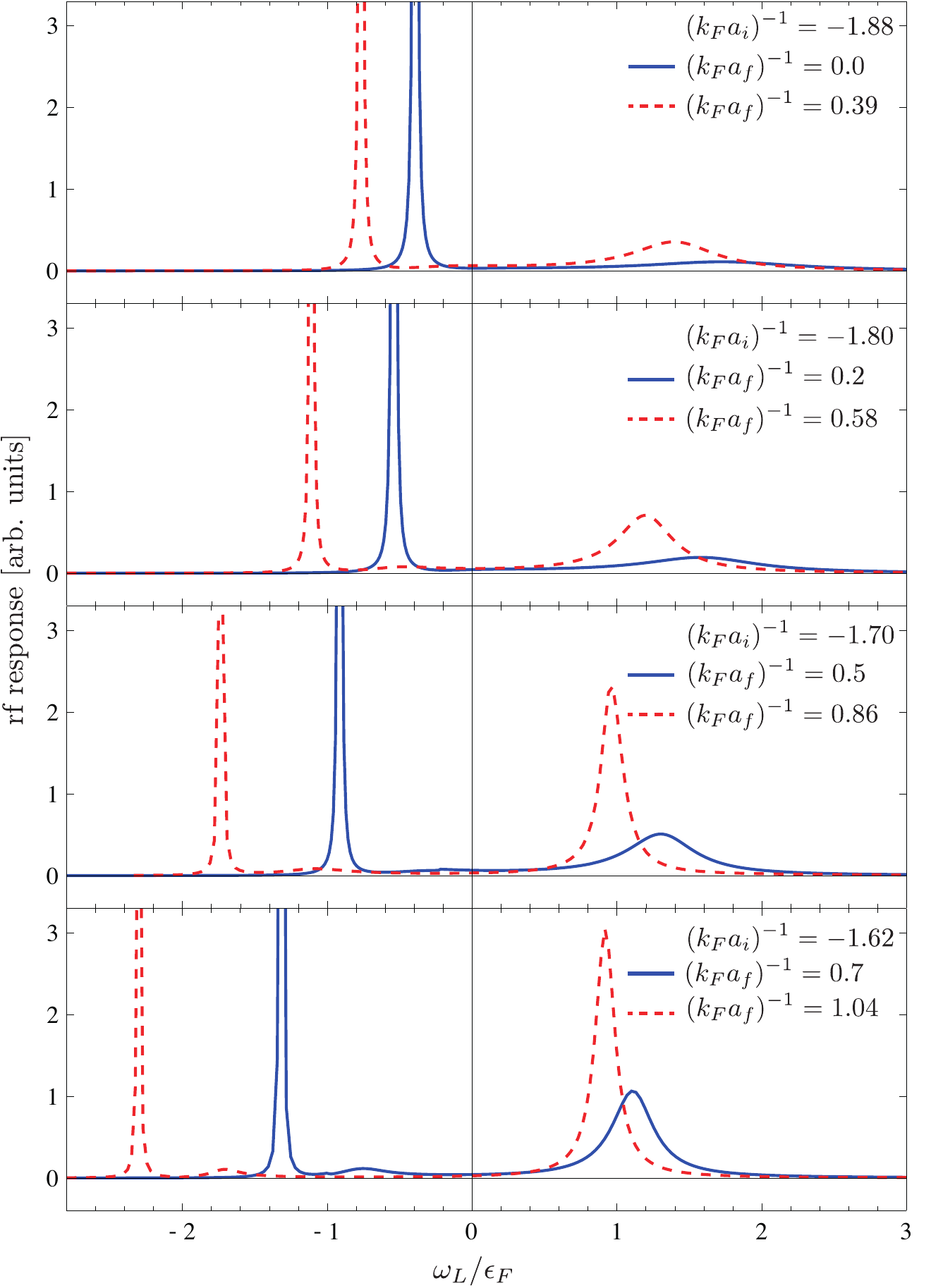}
  \caption{(color online).  rf-spectra $I(\omega_L)$ for minority
    species $\ket1$ (solid blue lines) and minority species $\ket3$
    (dashed red lines).  The interaction parameters correspond to the
    four magnetic field values marked in Fig.~\ref{fig:scattprofile},
    and listed in Table~\ref{tab:trans}: (a) top---(d) bottom.}
  \label{fig:rfspectra}
\end{figure}

We have calculated the parameters $m_\downarrow^*$, $Z_i$, and
$\Delta$ as functions of $(k_F a)^{-1}$ via the non-feedback
(non-self-consistent T-matrix) calculation.  $\mu_i$ is determined
self-consistently to ensure the correct impurity density.  Inserting
the spectral function Eq.~\eqref{initialA} into the susceptibility
\eqref{losusc} we obtain for the rf-response
\begin{multline}
  I(\omega_L)=\frac{\Omega^2Z_i}{2\pi^2}
  \int^{\sqrt{2m_\downarrow^*\Delta}}_0 dk
  \vec k^2 \\
  \times A_f\Bigl(\vec k, \mu_i-\mu_f+\omega_L-\Delta
  +\frac{\vec k^2}{2m_\downarrow^*}\Bigr).
  \label{eq:momsum}
\end{multline}
For the final state $\ket{f}$ we use the spectral function obtained in
section~\ref{sec:spectrum} from the full fRG calculation.  In
Fig.~\ref{fig:specmom} we show an example of the full final-state
spectral function in dependence of frequency and momentum which enters
the momentum sum in Eq.~\eqref{eq:momsum}.  One can clearly discern
the broadening of the attractive polaron branch at larger momenta, as
well as the broad repulsive branch at higher frequencies.  A similar
feature appears in the spectral function for the balanced Fermi gas
above $T_c$ as calculated by Haussmann \textit{et al.}
\cite{haussmann2009}.  The fermionic spectral functions
and rf-spectra in this work were determined using state-of-the-art
\textit{self-consistent} T-matrix (2PI) approximations and form the
basis for a full linear response calculation including vertex
corrections \cite{enss2011}.

Initially, the gas is prepared in a $\ket{1}$ and $\ket{3}$ mixture,
and both states can serve as minority or majority species, for
example, $\ket{\uparrow}=\ket{3}$ and $\ket{\downarrow}=\ket{1}$ such
that $\ket{i}=\ket{1}$ and $\ket{f}=\ket{2}$. The initial occupation
of the minority $\ket{\downarrow}$ states is small and for our
numerical calculation we use $x=n_\downarrow/n_\uparrow=0.01$. The
energy scale is set by the Fermi momentum of the majority species
$k_{F\uparrow} = 0.00015\,a_0^{-1}$ as appropriate for the MIT
experiment \cite{schirotzek2009}.  We calculate the rf-spectra at four
values of the magnetic field indicated as stars in
Fig.~\ref{fig:scattprofile}, and listed in Table~\ref{tab:trans}.  The
resulting spectra are shown in Fig.~\ref{fig:rfspectra}: the solid
blue lines indicate the response for minority species $\ket1$, while
the dashed red lines correspond to minority species $\ket3$.  In the
latter case the sign of the frequency offset $\omega_L$ is changed
because the state $\ket3$ is energetically above the final state.

The position of the sharp attractive polaron peak at negative
frequency offset $\omega_L$ shifts with the interaction parameter
$(k_Fa_f)^{-1}$ in accordance with the energy spectrum
Fig.~\ref{fig:energyspectrum}.  One observes that the attractive
polaron loses quasi-particle weight on the molecular side, cf.\
Fig.~\ref{fig:Zpol}.  In contrast, the repulsive polaron branch gains
quasi-particle weight toward the molecular side, and the respective
peak in the rf-spectra becomes both larger and narrower, and one can
read of the increasing lifetime.

The spectra in Fig.~\ref{fig:rfspectra} are convolved with a sinc
function $\sinc^2(\omega T/2)$ which gives the response to an rf-pulse
with a rectangular profile of length $T=20\,$ms \cite{haussmann2009}.
While our curves are computed for zero temperature, a finite
temperature $\sim 0.01\,T_F$ would lead only to a slight broadening of
the experimental rf-peaks. The broadening of the attractive polaron
due to the finite lifetime on the molecular side $(k_Fa_f)^{-1} >
(k_Fa_c)^{-1}$, however, is too small to be resolved. Note that we
have included both final-state and initial-state interaction in our
calculation: the knowledge of the spectral function for the initial
state allows for a detailed study of the final-state spectral
function.


\section{Discussion}
\label{sec:concl}

We have presented a new computational method to solve the
non-perturbative, exact renormalization group equation \eqref{wetteq}
and have demonstrated its efficiency for the Fermi polaron problem as
a specific example.  The inclusion of the full frequency and momentum
dependence of the propagators opens up new perspectives to apply the
functional renormalization group to problems where the detailed
dynamics of the relevant degrees of freedom becomes important
\cite{honerkamp2007}.  In particular, the method draws its strength
from the possibility to successively bosonize further channels of the
interaction via additional auxiliary fields (Hubbard-Stratonovich
transformation) \cite{husemann2009}.  In this way one can partially
capture the complicated analytical structure of higher-order vertex
functions $\Gamma^{(n)}$, including possible quasi-particle poles and
branch cuts, as we have explicitely shown for the $s$-wave scattering
channel in the polaron problem.  In combination with the recently
developed flowing rebosonization technique \cite{gies2002} our
numerical method can be extended to also incorporate re-emerging
vertices.  Our approach complements the proposal for bosons
\cite{blaizot2006} and additionally includes fermions.

For the Fermi polaron problem we achieve a unified description of many
dynamical effects beyond thermodynamics.  We verify the non-trivial
power-law scaling of decay rates \cite{bruun2010} and determine the
properties of the repulsive polaron with a method beyond the
non-self-consistent T-matrix approximation \cite{cui2010,
  massignan2011}.  This is of value in the ongoing debate about the
possible occurrence of ferromagnetism in ultracold Fermi gases with
short-range interactions \cite{jo2009, conduit2009, cui2010,
  pilati2010, pekker2011, barth2011, massignan2011}.  The polaron
problem sheds 
light on this question in the limit of strong population imbalance.
For the repulsive polaron we find the critical interactions strength
$k_Fa=1.57$ from our numerical data (Fig.~\ref{fig:energyspectrum}).
Going to a finite density of $\downarrow$-fermions is straightforward
within the fRG and involves only a slight modification of the
regulator of the $\downarrow$-fermions \eqref{sharpregulatorprops} as
long as no spontaneous symmetry breaking occurs.  By continuity we can
infer that the repulsive branch will remain to exist for small but
finite $\downarrow$-population and will exceed the critical energy
$\epsilon_F$ for the presumed onset of saturated ferromagnetism.  It
is an open question whether for larger impurity concentrations the
repulsive branch is so strongly renormalized that saturated
ferromagnetism can be ruled out \cite{barth2011}, or whether
competition with molecule formation may preclude the observation of
ferromagnetic domains \cite{pekker2011}.  Answering these questions
will require a full non-equilibrium calculation.

There has been much theoretical progress on the repulsive Fermi gas
with short-range interactions, but relatively few experiments have
been completed.  While the repulsive $^3$He Fermi gas has been studied
extensively in experiment, it is not dilute and has a large repulsive
hard-core potential \cite{woelfle1990}.  In contrast, ultracold Fermi
gases offer the realization of a proper contact interaction of tunable
strength.  We predict rf-transition rates for the repulsive branch and
propose a possible route to measure these excited states in a $^6$Li
Fermi gas.  This is a challenging problem because the final repulsive
polaron state is highly unstable.  One possible approach could involve
fast tomographic imaging similar to the MIT experiment
\cite{schirotzek2009}, another might be to measure the loss in the
final state which is expected to scale with the rf-transition rate for
a constant rf-pulse time but may be suppressed by the quantum Zeno
effect \cite{lompe2010}.  Hence, the possible observation of the
repulsive polaron represents not only a test of theoretical
predictions but poses an interesting challenge touching several
aspects of many-body physics.

\acknowledgments

We wish to thank M.~Barth, N.~Dupuis, T.~Hyodo, S.~Jochim, N.~Kaiser,
S.~Moroz, M.~Punk, S.~Rath, and W.~Zwerger for many useful discussions
and P.~Julienne, N.~Prokof'ev, B.~Svistunov, and M.~Zwierlein for
kindly providing their data for comparison.  Part of this work was
supported by the DFG within the Forschergruppe 801.


\appendix

\section{Derivative expansion}
\label{sec:derivative}

In this appendix we qualitatively study the polaron-to-molecule
transition using a simple approximation for the renormalized Green's
functions $G_{\downarrow,k}$ and $G_{\phi,k}$.  As we have seen,
$G_{\downarrow,k}$ and $G_{\phi,k}$ generally develop a complicated
frequency and momentum ($\omega,\mathbf p$) dependence. Here, we will
use an expansion in small frequencies and momenta (derivative or
gradient expansion) which allows for an analytical evaluation of the
loop integrals on the right-hand side of the flow equation
\eqref{genflow}.  The derivative expansion proves to be a good
approximation if one is interested in physics determined by the
structure of the Green's functions close to their poles as, for
example, in the description of phase transitions and critical
phenomena \cite{berges2002}.  In the derivative expansion the
dependence of the inverse propagators
$P_{\downarrow/\phi,k}(\omega,\p)$ on the RG scale $k$ is approximated
by
\begin{align}
  \label{derivativeprops}
  P_{\downarrow,k}(\omega,\p)
  &=A_{\downarrow,k}\cdot(-i\omega+\p^2)+m_{\downarrow, k}^2\\
  P_{\phi,k}(\omega,\p)
  &=A_{\phi,k}\cdot(-i\omega+\p^2/2)+m_{\phi, k}^2.\notag
\end{align}
In this approximation, which is similar to the one used in the
Wilsonian RG approach to the polaron problem \cite{gubbels2008}, we
assume that the renormalization of the frequency and momentum
coefficients is given by common wave-function renormalizations
$A_{\downarrow,k}$ and $A_{\phi,k}$, respectively, which in turn are
related to the quasi-particle weights via $Z_{\downarrow/\phi} =
A^{-1}_{\downarrow/\phi}$.  The flowing gap terms $m_{\downarrow,k}^2$
and $m_{\phi,k}^2$ are related to the flowing static self-energy via
Eq.~\eqref{massterm}.

From equation \eqref{genflow} one can derive the flow equations for
the four running couplings $A_{\downarrow,k}$, $A_{\phi,k}$,
$m_{\downarrow,k}^2$, and $m_{\phi,k}^2$.  The $\uparrow$-fermions are
not renormalized, $A_{\uparrow,k}=1$ and $m_{\uparrow,k}^2 =
-\mu_\uparrow$.  With the sharp cutoff \eqref{sharpregulatorprops},
the frequency as well as the momentum integrations can be performed
analytically.  The resulting flow equations read
\begin{align}
  \label{derivativeDGLsharp}
  \partial_k A_\downarrow
  &=-\frac{2h^2 k}{\pi^2 A_\phi}\theta(\mu_\uparrow-2k^2)
  \Bigl[\frac{\sqrt{\mu_\uparrow-k^2}}
  {(k^2+\mu_\uparrow+2 m_\phi^2/A_\phi)^2}\notag\\
  &\qquad+\frac{k}{(-k^2+2 \mu_\uparrow+2 m_\phi^2/A_\phi)^2}\Bigr]\\
  \partial_k m^2_\downarrow
  &=\frac{h^2 k}{\pi^2 A_\phi}\theta(\mu_\uparrow-2k^2)
  \Bigl[\frac{\sqrt{\mu_\uparrow-k^2}}
  {k^2+\mu_\uparrow+2 m_\phi^2/A_\phi}\notag\\
  &\qquad+\frac{k}{-k^2+2 \mu_\uparrow+2 m_\phi^2/A_\phi}\Bigr]\notag\\
  \partial_k A_\phi
  &=-\frac{h^2 k }{2\pi^2 A_\downarrow}\,
  \frac{\sqrt{\mu_\uparrow+k^2}}
  {(2 k^2+\mu_\uparrow+m_\downarrow^2/A_\downarrow)^2}\notag\\
  \partial_k m_\phi^2
  &=\frac{h^2 k}{2\pi^2 A_\downarrow}\,
  \frac{\sqrt{\mu_\uparrow+k^2}}
  {(2 k^2+\mu_\uparrow+m_\downarrow^2/A_\downarrow)}\notag.
\end{align}
The initial conditions for this system of differential equations
\eqref{derivativeDGLsharp} are specified at the UV scale $k=\Lambda$.
We note that the derivative expansion \eqref{derivativeprops} of the
molecule propagator $P_{\phi,k}(\omega,\p)$ cannot account for the
correct vacuum scattering amplitude \eqref{scattamp}, because the term
$iq$ is rooted in the non-analytical structure of the molecule
propagator \eqref{vacprop}.  We therefore focus only on the correct
calculation of the scattering length $a$ for $q=0$, which leads to the
infrared condition $(m_{\phi,k=0}^\text{vac})^2 = -h^2/(8\pi a)$ for
$\mu_\downarrow=\mu_\uparrow=0$.  In this case of two-body physics
$P_{\downarrow,k}$ is not renormalized, $A_{\downarrow,k}=1$ and
$m_{\downarrow,k}^2=0$, and the differential equations for
$A_{\phi,k}$ and $m_{\phi,k}^2$ decouple and can be solved
analytically.  The integration of \eqref{derivativeDGLsharp} in the
vacuum limit yields
\begin{equation}
  (m_{\phi,k=0}^\text{vac})^2
  = m_{\phi,\Lambda}^2-\frac{h^2\Lambda}{4\pi^2}.
  \label{mphik}
\end{equation}
This leads to the UV condition for the molecule gap,
\begin{equation}
  m_{\phi,\Lambda}^2=\frac{h^2}{8\pi}(2 \Lambda/\pi-a^{-1}),
\end{equation}
which incorporates the correct regularization of the UV divergence
$\Lambda$ in Eq.~\eqref{mphik}.  At the UV scale the momentum and
frequency dependence of $P_{\phi,\Lambda}$ can be neglected due to the
large bosonic gap $m_{\phi,\Lambda}^2$, and we set
$A_{\phi,\Lambda}=1$.

For a finite density of $\uparrow$-atoms the system of differential
equations \eqref{derivativeDGLsharp} is solved numerically.  The
initial values for the $\downarrow$ propagator are then given by
$m_{\downarrow,\Lambda}^2=-\mu_\downarrow$ and
$A_{\downarrow,\Lambda}=1$.  The down chemical potential
$\mu_\downarrow$ is determined in the way discussed in
section~\ref{sec:frg}. We find that the polaron is indeed the ground
state for interaction strengths $(k_Fa)^{-1} < (k_Fa_c)^{-1}$ whereas
the molecule becomes the ground state for $(k_Fa)^{-1} >
(k_Fa_c)^{-1}$, with $(k_Fa_c)^{-1}=0.96$.

The energy spectrum from the simple derivative expansion is in
qualitative agreement and even in rough quantitative agreement with
the results obtained from our new numerical method and other
theoretical calculations.  A drawback of the derivative expansion is
that it is impossible to extract a reasonable spectral function from
an ansatz of the form \eqref{derivativeprops} as it only accounts for
a single coherent quasi-particle excitation.  Neither higher excited
states, such as the repulsive polaron, nor the incoherent background,
which comprises the major weight for the molecule (cf.\
section~\ref{sec:spectrum}), can be captured with this ansatz.
Furthermore, although decay processes lead to a finite lifetime of the
excited polaron and molecule branches, in the simple approximation
\eqref{derivativeprops} these states have a vanishing decay width.
Note that the finite lifetime of excited states can be obtained
neither from simple variational wave functions \cite{chevy2006} nor
from the non-self-consistent T-matrix approach \cite{combescot2007}.
Their description requires the full self-energy feedback developed in
section~\ref{sec:solution}.



\end{document}